\documentclass[1p,times,onecolumn]{elsarticle}

\usepackage{booktabs}
\usepackage{hyperref}

\usepackage{pdflscape}
\usepackage{longtable}
\usepackage{subfigure}
\journal{Journal -- TBD}

\biboptions{authoryear}   

\usepackage[nottoc]{tocbibind}

\usepackage{array}

\begin{document}
\begin{frontmatter}

\title{Exposure of occupations to technologies of the fourth industrial revolution}

\author[add1]{Benjamin Meindl}
\author[add2,add3,add4]{Morgan R. Frank}
\author[add1,add5]{Joana Mendonça}

\address[add1]{Center for Innovation, Technology and Policy Research (IN+), Instituto Superior Técnico, Universidade de Lisboa}

\address[add2]{Department of Informatics and Networked Systems, University of Pittsburgh}
\address[add3]{Media Laboratory, Massachusetts Institute of Technology}
\address[add4]{Digital Economy Lab, Stanford Institute for Human-Centered Artificial Intelligence, Stanford University}
\address[add5]{CEiiA}

\begin{abstract}

The fourth industrial revolution (4IR) is likely to have a substantial impact on the economy. Companies need to build up capabilities to implement new technologies, and automation may make some occupations obsolete.  
However, where, when, and how the change will happen remain to be determined. 
Robust empirical indicators of technological progress linked to occupations can help to illuminate this change. With this aim, we provide such an indicator based on patent data.
Using natural language processing, we calculate patent exposure scores for more than 900 occupations, which represent the technological progress related to them.
To provide a lens on the impact of the 4IR, we differentiate between traditional and 4IR patent exposure. Our method differs from previous approaches in that it both accounts for the diversity of task-level patent exposures within an occupation and reflects work activities more accurately. 
We find that exposure to 4IR patents differs from traditional patent exposure.
Manual tasks, and accordingly occupations such as construction and production, are exposed mainly to traditional (non-4IR) patents but have low exposure to 4IR patents. 
The analysis suggests that 4IR technologies may have a negative impact on job growth; this impact appears 10 to 20 years after patent filing.
Researchers could validate our findings through further analyses with micro data, and our dataset can serve as a source for more complex labor market analyses.
Further, we compared the 4IR exposure to other automation and AI exposure scores. Whereas many measures refer to theoretical automation potential, our patent-based indicator reflects actual technology diffusion.
We show that a combination of 4IR exposure with other automation measures may provide additional insights. For example, near-term automation might be driven by non-4IR technologies.
Our work not only allows analyses of the impact of 4IR technologies as a whole, but also provides exposure scores for more than 300 technology fields, such as AI and smart office technologies. Finally, the work provides a general mapping of patents to tasks and occupations, which enables future researchers to construct individual exposure measures.

\end{abstract}
        

\end{frontmatter}

\section{Introduction}\label{introduction}

Technological progress continuously impacts the economic environment. The current wave of technological progress is driven by digitalization and the adoption of artificial intelligence (AI), and is often referred to as the fourth industrial revolution. AI might enable machines to become increasingly able to perform tasks that previously only humans could perform. Whereas previously machines were mainly able to perform clearly defined, repetitive, routine tasks \citep{Acemoglu2011}, future automation might cover much more diverse tasks, for example, some requiring emotional intelligence \citep{Brynjolfsson2017_m}. These new automation patterns create fears of machines making workers obsolete and creating unemployment. However, technological change has various effects on the labor market, and previous waves of automation did not result in long-lasting technology-induced rising unemployment \citep{Mokyr2015,Autor2015,JamesB2019}.

\cite{Acemoglu2019} observe that automation not only decreases labor demand but also has a productivity effect, which increases labor demand. Further, they describe the reinstatement effect, where automation leads to newly-created tasks carried out by humans. The relative size of these effects and their interaction determine the overall effect of automation on the labor market. Therefore, automation causes changes in the task content of occupations due to lower demand for some tasks, higher demand for remaining tasks, and the creation of new tasks.

To evaluate these effects and prepare for future shifts in the labor market, researchers, e.g., \cite{Frey2017} and \cite{Brynjolfsson2017_m}, construct measures of automation potential of occupations. These measures can provide valuable insights for future research in terms of overall automation potential. Our approach does not aspire to predict the share of automated jobs, but aims to reflect actual technology maturity (diffusion), which is not covered by the aforementioned indicators \cite{Arntz2019}. For example, scores by \cite{Brynjolfsson2017_m} are based on expert assessments of ``what can machine learning do?'' We use patent data as an indicator for technological progress; patents actually document what existing technology can currently do. The McKinsey Global Institute follows a similar objective and focuses on what actual automation might look like until 2030, acknowledging that there is a much higher automation potential in the long term \citep{McKinsey2017}. They provide estimates of automation potentials per occupation, which they expect to be implemented until 2030.

Linking patent data to occupation activities offers a direct indicator of the exposure of occupations to technology.  There exist patent occupation mappings at an industry \citep{Silverman2002} and occupation level \citep{Kogan2020, Webb2020} which have been used for economic analyses \citep{Mann2017, Acemoglu2020c}. \cite{Webb2020} found, for example, that exposure to previous automation technologies had a negative impact on employment at an occupation level, and \cite{Mann2017} identified an overall positive impact of automation patents on employment.  

We build on the approach of \cite{Kogan2020} and refine for improved accuracy and to account for task-level differences. Each occupation relates to several tasks, and technology exposure may vary among different tasks within an occupation \citep{Brynjolfsson2017_m}. The task level, as the ``unit of work that produces output,'' is a highly insightful level of detail for evaluating the impact of technologies on jobs \citep{Acemoglu2011}. Our approach has two main benefits. On the one hand, it allows accounting for a specific technology exposure for each task, which is ignored when looking at occupations as a whole. On the other hand, our task-level approach increases the accuracy of the mapping, as it identifies patents specific to each activity, rather than patents which have many words in common with the overall occupation description. For example, our approach might avoid associating a robot engineer mainly with robot patents in general (e.g., improved efficiency of assembly robots), but rather with patents which describe innovations that help to ``plan robot path”, ``debug robot programs”, and ``maintain robots.”
Further, we introduce a measure of technology exposure; we therefore differentiate between technologies of the fourth industrial revolution (4IR patents) and other patents (non-4IR patents) for creating technology exposure scores.
These scores indicate patent exposure at the task and occupation level. Our analysis includes patent data since 1970 and thus allows us to review developments over time, e.g., when 4IR technologies have been developed and how long it takes them to impact the labor market.

Various researchers identify the lack of high-quality data on technological progress of key 4IR technologies as a key barrier to better understanding the impact of those technologies on the workforce \citep{Frank2019c,Mitchell2017}. With this article, we address this issue by providing a mapping of patents to occupational tasks and introducing a 4IR technology exposure score per occupation.

\section{Patents as an indicator for technological change}\label{sec:background}

    Several studies build on patent data as an indicator of technological progress or innovative activity. \cite{Silverman2002}, for example, create a concordance table of technologies used and produced per industry to analyze the impact of technological resources on corporate diversification. The work builds on manually-annotated patent data from the Canadian patent office. \cite{Mann2017}, for example, build on this dataset, to evaluate the effects of automation technologies per sector of use. The dataset is generally interesting but builds on data from 1990–1993 and only provides a patent category-to-industry mapping. Therefore, it is not suitable for a fine-grained mapping of patents to occupations, particularly for newer technologies.

	Additionally, \cite{Dechezlepretre2019} use patent data to evaluate the relation of wages and automation innovations. They focus on advanced manufacturing patents, as defined by \cite{Aschhoff2010} combined with their own patent search terms.  They link those patents to industry sectors according to a concordance table provided by \cite{Lybbert2014}. This patent-industry mapping is based on industry-specific terms. The mapping does not account for activities conducted by workers within industries; for example, industry descriptions contain terms related to the output, e.g., ``manufacture cars,'' and not the task, such as ``cut sheet metal.'' 

	\cite{Webb2020} proposed an approach to link patent data more directly to occupations. He extracts verb-noun pairs from patent titles and task descriptions, and uses those as a basis for mapping. For example, if a doctor's task is ``diagnose a patient’s condition,'' the verb-noun pair is ``diagnose condition.'' The analysis thus aims to identify patents describing similar actions as task descriptions. Similar tasks can be described with different words, e.g., ``diagnose condition'' may be very similar to ``diagnose disease.'' Webb overcomes this challenge by using word hierarchies to identify more general words for the matching (higher-level terms per word). The previous examples would both become ``diagnose state.''
	
	\cite{Kogan2020} also uses natural language processing for calculating similarity scores of patents and occupation descriptions. Instead of relying on word hierarchy information, they use text embeddings. They represent words through vectors, which are trained on large amounts of text data. Words that are more similar are represented as vectors that are more similar. For example, ``king'' minus ``man'' plus ``woman'' leads to a vector similar to ``queen.''  Word vectors allow the calculation of similarity scores of words, and a more fine-grained differentiation is possible. Further, the approach does not rely only on verb-noun pairs (such as \citealt{Webb2020}), but on the full-text data. This enables the approach to account better for context descriptions. For example, it differentiates between ``diagnose \textit{patients} condition'' and ``diagnose \textit{machine} condition.''
	
	We aim to build on this general idea of mapping patent data to occupations through text embedding, while implementing the mapping at a task level. This has two key advantages.
	First, \cite{Kogan2020} use task descriptions for an occupation and combine them into one text. They compare this block of text with patent texts to identify the most relevant patents. However, technology exposure may vary for different tasks within an occupation \citep{Brynjolfsson2017_m}.  Therefore, the task level, as the ``unit of work that produces output,'' is a highly insightful level of detail for evaluating the impact of technologies on jobs \citep{Acemoglu2011}. For example, a robot technician does ``attach wires between controllers,'' and ``develop robotic path motions.'' The tasks are very different and probably have exposure to different technologies.
	Second, the occupation-level approach has a chance of identifying general context patents, ignoring the broad range of patents related to the different activities of an occupation. For example, a barber performs tasks such as ``clean and sterilize scissors,'' ``recommend and sell lotions,'' and ``shampoo hair.'' Stringing together all task descriptions of a barber will lead to a text with many words such as shampoo, hair, and lotion. Therefore, it is likely that there is a high similarity score to patents describing hair care products. Looking at task statements independently accounts better for words describing the actual activity and identifies patents related to the activity itself, e.g., \textit{sell} lotions. Further, if there are many tasks related to general activities, such as working with hair, and few tasks related to secondary activities, such as bookkeeping, the occupation level search is likely to ignore the secondary activities.

\section{Mapping patents to occupations}\label{Mapping}

\begin{figure}[h!]
    \centering
    \includegraphics[width=0.6\textwidth]{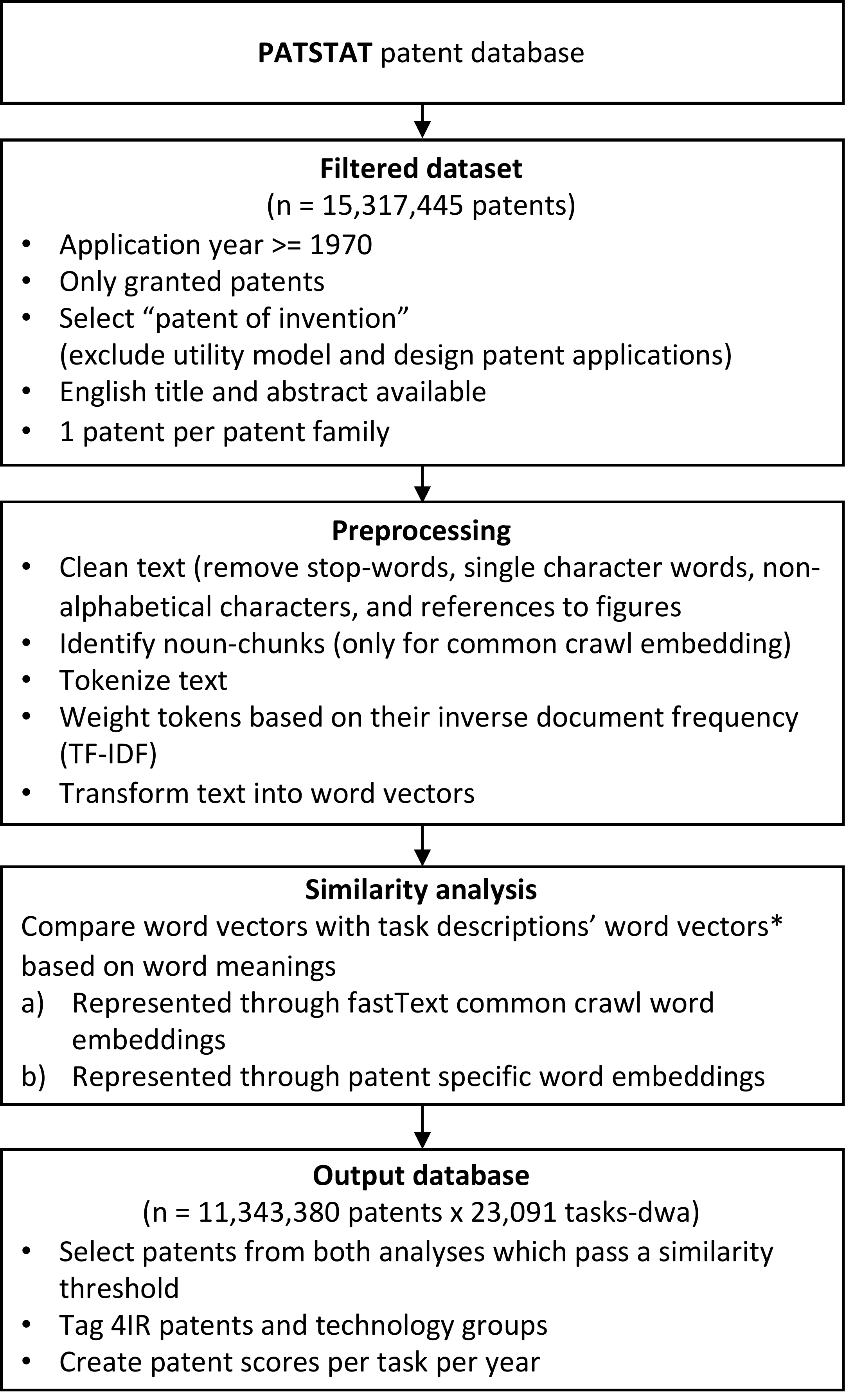}
    \caption[Description of our analysis to identify relevant patents per task.]{Description of our analysis to identify relevant patents per task. This chapter provides a more detailed description of the processing steps. \newline
    \** We conducted a similar preprocessing for task descriptions.}
    \label{fig:flowchart}
\end{figure}

To map occupations to patents, we compare task descriptions with patent abstracts. Therefore, we use natural language processing methods. Figure \ref{fig:flowchart} provides an overview of the approach.

\subsection{Data}
We compare patent texts with descriptions of activities conducted in an occupation. We build on occupation descriptions from O*NET, which describes each occupation as 20 to 40 tasks (more than 20k tasks). This approach helps to identify the most relevant patents per occupation and occupational task. All task descriptions contain information about the activity conducted. Also, some task descriptions contain information about the tools and technologies used. We focus on activities only, and thus remove references to tools used. For example, for the task ``Monitor geothermal operations, using programmable logic controllers,'' we would remove ``using programmable logic controllers.'' This ensures that the search includes all patents related to monitoring geothermal operations, independently of the technology they use.

Our analysis builds on the PATSTAT patent dataset. We include all patents since 1970 with an English abstract in our analysis. 
Identical patents registered in multiple patent offices share a common patent family ID. We only select one patent per family to avoid double counting. For each patent family, we choose patents in the following order: preferably we use USPTO patents and those of other English-speaking countries' patent offices, followed by WIPO and EPO patents. 
For the task similarity scores we rely on patent abstracts. Abstracts are highly suitable for this task and comprise the densest information on inventions \citeauthor{Benzineb2011}.
Our dataset includes 15,317,445 patents filed between 1970 and 2020. Figure \ref{fig:appendix_patstat_per_year} shows that the number of patents increased until 2013, then sharply decreased, particularly after 2017. The recent decline is due to the time lag between the filing date and publication in the PATSTAT database. Further, the total number of patents decreased in 1976, 1991-1992, and in 2004. Due to the substantial decline in patents after 2015, our analysis offers its most valuable insights from 1970 to 2015.

\subsection{Patent task similarity scores}\label{sec:docsim}

We identify the most relevant patents per task by comparing the texts of task descriptions and patent abstracts. Our algorithm builds on word embeddings, where multidimensional word vectors represent words. Vector representation identifies text similarity, even if there are no words in common. A similar approach has been used by \cite{Kogan2020}, who map patent texts to occupation descriptions.
The approach yields the best results when the queries comprise at least a few sentences. For very short queries (such as task statements), there is a risk that single words bias the results. For example, the task ``Develop engineering specifications or cost estimates for automotive design concepts'' describes the activity (cost estimates) and the context (automotive design). There is a risk that results may include patents only related to the context, if there is a high word overlap, e.g., a patent describing ``cost-efficient automotive design.'' We try to avoid a bias towards the context, and combine task statements that describe similar activities for our queries. Therefore, we build the O*Net dataset, which clusters task statements into 2067 ``detailed work activities'' (DWAs), such as ``estimate operational costs.'' The DWA ``estimate operational costs'' includes tasks: the tasks mentioned above, for example,  as well as the task ``analyze, estimate, or report production costs.'' On the one hand, combining both statements increases the weight of the activity (cost estimation), as this is common in all task statements. On the other hand, combining these statements reduces the weight of the context (automotive design), as many tasks describe a slightly different context. The different context words are valuable for narrowing the broader scope of the activity. In this case, the context relates to (cost estimation in) a production-related setting and, e.g., is not related to software development or services. For our queries, we combine all tasks related to one DWA. We create a unique query for each task by over-sampling its task statement and thus giving higher weights to the actual task description.

In order to work with the text data, we conduct some preprocessing steps. We remove expressions in brackets from patent abstracts, such as numeric references or abbreviations, which do not add relevant information for comparison with task descriptions. Further, we remove non-alphanumeric characters and single-character expressions. Also, we remove stop words, which are frequent words, such as ``is'' or ``and'', and which do not add much information to the text. Using only the most relevant words per text is commonly performed to increase accuracy when working with text. Finally, we transform all characters to lower case.

We transform text data for analysis in three steps.
First, we count word frequencies per document and weigh the frequency with the word frequencies across all texts. This method is called text frequency-inverse document frequency (TF-IDF). TF-IDF assigns higher weights to infrequent words, emphasizing those words that are most specific for a document.
Second, we create a word similarity matrix using word embeddings. 
Word embeddings are created based on large corpora, where algorithms assign vectors based on word relations to neighboring words. Those word vectors represent the meaning of words and allow for comparisons and calculation of text, without relying on the presence of similar words.
In our example, word vectors enable the identification of a similarity, e.g., for ``steer a car'' and ``drive the vehicle,'' even though there are no common words.
Third, we calculate similarity scores for each patent task combination using TF-IDF scores and the word similarity matrix. This leads to a task-patent similarity matrix of 23,091 task-DWA combinations and 15,317,445 patents, with similarity scores ranging from 0 to 1, with 1 representing similar documents. We reduce the matrix size for faster processing by setting irrelevant scores below a certain threshold (see below) to zero. 

We conduct this approach with two different word embedding algorithms to account for patent-specific terms and general language, and identify a high number of relevant documents. 
On the one hand, we use fastText word embedding, which has been trained on 600 billion tokens (words) from the common crawl corpus\footnote{\url{https://commoncrawl.org/}} \citep{Mikolov2018}. The embedding provides 300-dimensional vectors for around 2 million words, n-grams (combinations of words), and sub-words (parts of words, which can be used to construct vectors for unknown words).
The common crawl corpus comprises texts from various sources available on the web, and therefore the embeddings represent various aspects of language. However, the structure and words used in patent data are more technical than standard language. This leads to noise in our results when we compare task descriptions and patents. We only consider patents related to a task if the similarity score is above a certain threshold. 
The mean similarity score is 0.020, and many highly relevant patents are in the long tail of higher relevance scores. Manual review shows that above a threshold of 0.197, or 9 standard deviations above the mean, there is a high density of relevant patents. 
With this approach, we identify hundreds of relevant patents per task.

We repeat the approach with another word embedding, which has been trained on five million patents (over 38 billion tokens) \citep{Risch2019}. This embedding accounts for patent-specific language. It therefore helps to identify patents which use unusual words and would not have been identified though the common language approach. In turn, it has a higher noise caused by the language used in task descriptions, which is not necessarily similar to patent language. Similarly to the other embedding, the threshold value of nine standard deviations above the mean (0.170) proves to be a practical cutoff value to exclude most of the irrelevant patents.

On the one hand, combining both approaches allows us to sustain a high cutoff value and thus reduce noise through irrelevant patents. On the other hand, both searches complement each other and increase the overall number of patents mapped to tasks. We find that both approaches show identical overall results patterns. For example, both embeddings lead to a similar distribution of patents per task category (see \ref{appendix:validity} for detailed evaluation results). These identical patterns indicate that the approach is robust enough to ensure that the choice of word embedding did not systematically bias the results. Simultaneously, relying on both word embeddings significantly contributed to covering a broad range of patents and enabled the mapping of 75\% of patents in our sample to at least one task. 

We also conducted our analysis with cutoff values of ten and twelve standard deviations. Those additional analyses showed a similar evolution of number of patents per year and similar distribution patterns of total patent exposure across occupations. This suggests that the nine standard deviation cutoff value is sufficient, and higher cutoff values do not change the mapping significantly. Each standard deviation of higher cutoff value reduced the total coverage of patents associated with tasks by around 10 percent; we therefore choose to conduct our analysis with the nine standard deviation cutoff value.

\subsection{Patents of the fourth industrial revolution}  \label{sec:methods_4IR}

The current wave of digitization is often described as the fourth industrial revolution (4IR). \cite{HenningK2013} describe the 4IR as merging physical and virtual environments into cyber-physical systems. Since then, a large amount of research has evolved in this area. Several definitions exist and the literature frequently describes the Internet of Things as the core of the 4IR, as it enables smart technologies to transform traditional enterprises \citep{Meindl_Frank_2020}.
Similarly, \cite{Meniere2020} describe the 4IR as the ``full integration of information and communication technologies (ICT) in the context of manufacturing and application areas such as personal, home, vehicle, enterprise and infrastructure [...] towards a fully data-driven economy.'' \citeauthor{Meniere2020} build on this definition and leverage the expertise from the European Patent Office (EPO) to identify patents related to the 4IR across 350 4IR technology fields.

\citeauthor{Meniere2020} describe technology fields along two dimensions. First, they describe application domains, such as ``healthcare'' and ``industrial.'' Their second dimension describes technologies, such as ``software'', ``connectivity'', and ``core AI''\footnote{They identify three core technology fields, IT hardware, software, and connectivity, as well as eight enabling technology fields, including data management, user interfaces, core AI, geo-positioning, power supply, data security, safety, three-dimensional support systems. Application domains include consumer goods, home, vehicles, services, industrial, infrastructure, healthcare, and agriculture}.
For each of the fields, they identify multiple ranges of patent classification classes, describing these technologies. Each patent is assigned to at least one classification. We consider a patent to be a 4IR patent if any of its classifications falls within one of the 4IR classifications defined by \cite{Meniere2020}.

\subsection{Aggregation of results}

In the following sections, various analyses refer to the aggregated number of patents, e.g., average patents per task related to an occupation. When an aggregating task counts to activity or occupation counts, we use log values. This allows us to minimize the impact of outliers, i.e., tasks with a very high number of associated patents, and reflects whether there are generally  many tasks with high or low patent counts.
For aggregation by occupation cluster, we first aggregate at an occupation level and next on a cluster level to avoid bias through the number of tasks per occupation.
When aggregating task data, we weight the task scores with the task importance score for an occupation, as provided by O*Net, ranging from above 0 to 1. This leads to less important tasks having less impact on an occupation's exposure score. A similar approach has been followed by \cite{Brynjolfsson2017_m} when calculating SML scores for occupation.

\cite{Acemoglu2011} provide a classification of six task types (e.g., routine manual, cognitive analytical), based on work activity and work context variables, as described by O*Net. We apply their categorization in order to assign task types to each tasks. To do so, we link those variables to task descriptions as follows.
The work activity descriptions, e.g., ``4.A.2.a.4 Analyzing data/information'' map directly to detailed work activities and thus to task descriptions. The work context variables, e.g., ``4.C.3.b.7 Importance of repeating the same tasks'' map to both skills and abilities, both of which map to work activities. Therefore, we map work context variables via the intermediate variables skills and abilities to the task descriptions. We calculate z-scores per measure and aggregate those for each task category. For the analysis, we associate all tasks with a positive z-score to a category, which may lead to some tasks being associated with more than one category.

\section{Patent task mapping results}

The previous section identifies a mapping that shows the most relevant patents for each occupational task. Overall, our analysis associates 11,343,380 patents to tasks, thus 74\% of all patents in our dataset are associated with tasks. Not all technology clusters are equally present in our results. On the one hand, chemistry or biotechnology patents only map to few occupations, which is not surprising, as those inventions mostly do not describe activities, but rather formulas or materials. On the other hand, computer technology and IT methods for management patents are highly relevant for many tasks. 

To evaluate the quality of the patent-task mapping, we manually investigate the relevance of patents with a high similarity score. Table \ref{patent_task_examples} shows some example tasks and patents.

\begin{table*}[h]
\centering
\caption{Most similar patents per tasks.}
\label{patent_task_examples}
\begin{tabular}{p{0.1\textwidth}p{0.8\textwidth}}
\toprule
Similarity & Patent title \\
\midrule
\multicolumn{2}{p{0.9\textwidth}}{\textbf{Operate diagnostic or therapeutic medical instruments or equipment:} Assemble and use equipment, such as catheters, tracheotomy tubes, or oxygen suppliers.}  \\
\midrule
0.42	&	Fast trachea incising device	\\
0.40	&	Device and method for electrocardiography on freely moving animals	\\
0.38	&	Method and apparatus for weaning ventilator-dependent patients	\\
0.38	&	Method and apparatus for weaning ventilator-dependent patients	\\
0.37	&	Devices, systems and methods for using and monitoring tubes in body passageways	\\
\midrule
\multicolumn{2}{p{0.9\textwidth}}{\textbf{Prepare scientific or technical reports or presentations:} Write up or orally communicate research findings to the scientific community, producers, and the public. } \\
\midrule
0.52	&	Research product automatic register service method and system	\\
0.48	&	System and method for medical image interpretation	\\
0.47	&	Research collection and retention system	\\
0.46	&	Method and apparatus for the design and analysis of market research studies	\\
0.45	&	A System for Joint Research on the Internet	\\
\midrule
\multicolumn{2}{p{0.9\textwidth}}{\textbf{Evaluate quality of materials or products:} Perform visual inspections of finished products.} \\
\midrule
0.40	&	Mobile unit for express-control of oil products characteristics	\\
0.40	&	Automatic monitoring method for coated products and fabrication process	\\
0.38	&	Method and apparatus for inspecting manufactured products for defects in response to in-situ monitoring	\\
0.37	&	Automatic quality inspection method, device and system	\\
0.37	&	Optical fiber sensing system consistency test method	\\

\bottomrule
\end{tabular}
{
\begin{flushleft}
\small
Note: Multicolumn lines include detailed work activities (bold) and task descriptions. Our approach maps patents not only based on the actual task description, but includes information from detailed work activities. Two different word embedding methods have been used; the table includes the average similarity score of both embeddings.
\end{flushleft}
}
\end{table*}

The example results indicate a good match of task descriptions and patent results. Our analysis is conducted at a task level and also contains information on other tasks within the same detailed work activity (DWA). Therefore, the results may include patents not directly related to the task but capturing the activity described by the DWA.

\ref{appendix:clusters} provides additional information on the presence of patents per technology field and task type. 
In the following section, we review patent task mapping quality, task exposure scores, and evolution over time.

\subsection{Patents over time}\label{sec:pat_per_year}

\begin{figure}[h]
    \centering
    \includegraphics[width=0.6\textwidth]{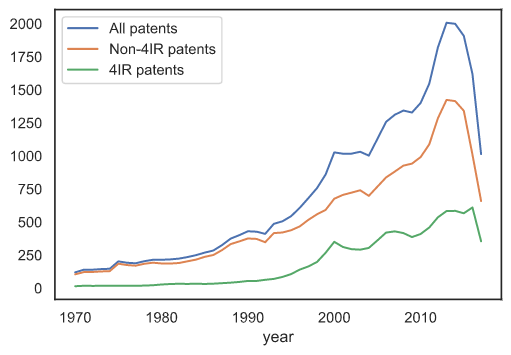}
    \caption[Average number of patents per task and year.]{Average number of patents per task and year. The figure shows a plateau between 2000 and 2004 after strong growth in the previous years. The plateau in overall patent exposure between 2000 and 2004 is driven by a decline in 4IR patents during this period.}
    \label{fig:mean_pat_per_year}
\end{figure}

Figure \ref{fig:mean_pat_per_year} shows the mean number of patents associated per task, per year. The mean number of patents per task increased over time, mostly in line with the overall number of patents in our sample. An exception is the period between 2000 and 2004, where the number of patents per task remained nearly constant, whereas the total number of patents in our raw data decreased only in 2004 and otherwise increased. Another plateau is in the years before 2008, with a substantial increase after that, whereas the curve of the total number of patents in our raw data is more smooth. Finally, the overall number of patents declined in recent years. This is because not all of these patent applications have yet been published in the PATSTAT dataset (see Figure \ref{fig:appendix_patstat_per_year} for total patents in our dataset).

\cite{Meniere2020} identify technologies of the fourth industrial revolution, which include, e.g., virtual reality, machine learning algorithms, and 3D modeling software. Some of those patents already existed decades ago, e.g., machine learning is a technology introduced decades ago, but which reached market maturity only recently. We compare overall patent exposure to 4IR patents with general exposure. Our findings show that exposure to Industry 4.0 patents grew more than overall exposure to patents. Our findings indicate that 4IR patents drove the strong growth in patents between 1990 and 2000. The share of Industry 4.0 patents of all patents in the results is 29\%, ranging from 12\% in 1976 to 38\% in 2016. A drop in Industry 4.0 patents after the year 2000 also explains the plateau of overall patents per task, whereas the number of non-Industry 4.0 patents continued its constant growth.
Interestingly, other research articles also observe this change in patenting behavior. \cite{Kelly2018}, for example, identify ``breakthrough patents'', which relate to novel technologies and differ in text content from previous patents. They observe a strong rise of breakthrough patents between 1980 and 2000, with a steep drop afterwards.

\subsection{Patents per task}

\begin{figure}[h]
\begin{center}
\subfigure[Distribution of number of tasks per patent.]{
    \includegraphics[width=0.48\textwidth]{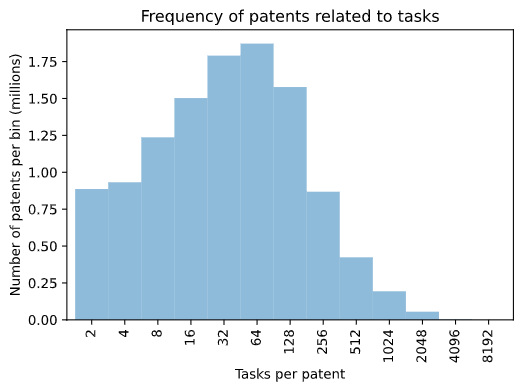}
    \label{fig:hist_task_patent}
    }
\subfigure[Distribution of number of patents per task.]{
    \includegraphics[width=0.48\textwidth]{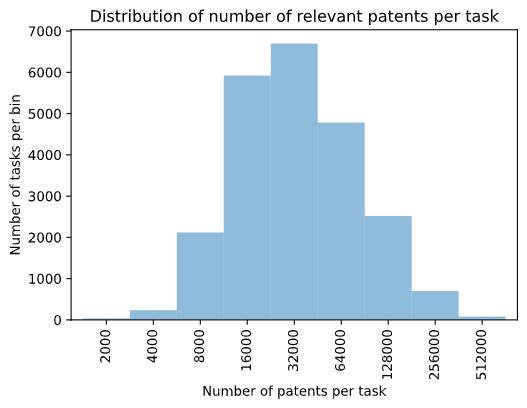}
    \label{fig:hist_sum_pat}
    }
\caption{Patent frequencies and number of patents per task.}
\label{fig:histograms_pat_task}
\end{center}
\end{figure}

Most of the 11 million patents in our results map to multiple tasks. On the one hand, task descriptions share similar content, and DWA-level information forms part of the search queries, which leads to a patent being likely to be mapped to multiple patents within a DWA. On the other hand, a patent may describe an invention which is relevant for different tasks. Figure \ref{fig:hist_task_patent} shows the frequency of patents in our results. The x-axis indicates how many tasks a patent is associated with (out of more than 23,000 tasks). Most patents relate to around 30 to 60 unique task/DWA combinations.

Next, we look into the differences of relevant patents per task. Figure \ref{fig:hist_sum_pat} shows that the number of patents per tasks ranges from around 2,000 to more than 500,000 patents. The distribution of the number of patents per tasks is log-normal, with most tasks associated to 32,000 patents. To account for the logarithmic distribution of patents per task, we use the logarithm of the number of patents for further analysis, where we aggregate numbers of patent per task, to avoid bias through tasks with high numbers of patents. Tasks with the lowest number of relevant patents are related to the activities ``prepare operational budgets for green energy or other green operations,'' ``collaborate with others to determine technical details of productions,'' and ``manage budgets for personal services operations.''
Tasks with the highest number of relevant patents are related to the activities ``thread wire or cable through ducts or conduits,'' ``record research or operational data,'' and ``prepare data for analysis.''

\subsection{Evaluation per task type}
\label{sec:tasktype}

\begin{figure}[h]
    \centering
    \includegraphics[width=1.1\textwidth]{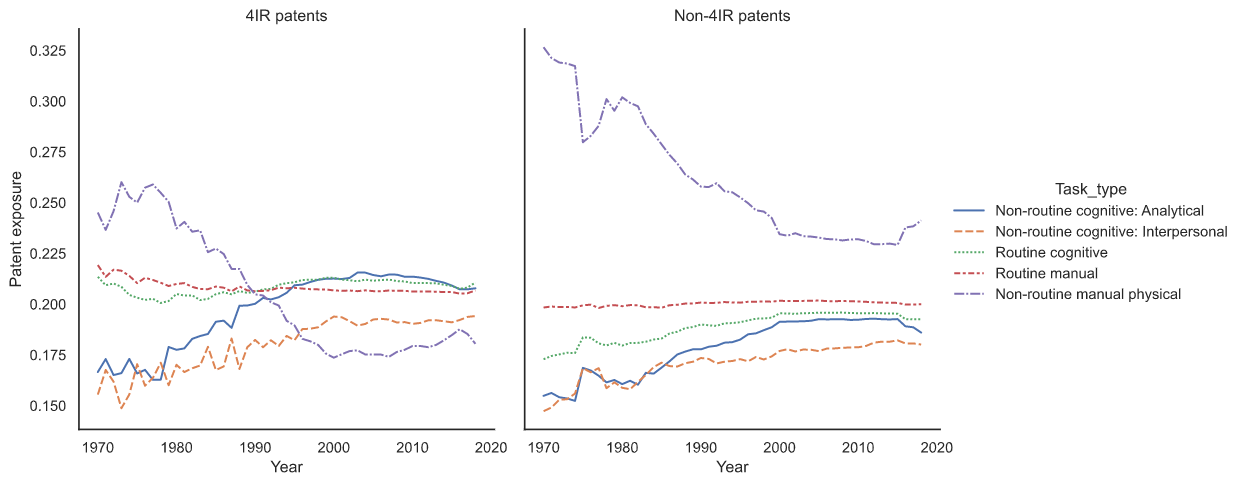}
    \caption[Number of patents per task type over time.]{Number of patents per task type over time. The graphs show total exposure per task and exposure to 4IR patents and non-4IR patents. The exposure values indicate the share of patents per year and task type.}
    \label{fig:pat_per_task_year}
\end{figure}

To better understand the patent exposure of tasks over time, we categorize tasks into six different task types (e.g., routine, physical, cognitive), based on the classification provided by \cite{Acemoglu2011}. 
Our analysis provides a patent count per task and year. For each task, we evaluate the number of patents over time. We aggregate the log number of patents per task type to see the overall evolution of patents per task type. \ref{fig:pat_per_task_year} shows the evolution of patents over time.

Section \ref{sec:pat_per_year} showed that there is a strong growth in the overall number of patents (for all task types); this section shows each task type’s share of total patent exposure per year. Non-routine manual tasks have the highest number of related patents, but with the rise of 4IR technologies their share of total patents decreased. They have the lowest exposure to 4IR patents. In contrast, analytical tasks experienced a strong growth in patent exposure, driven by a high share of 4IR patents. This is in line with the exposure to breakthrough technologies, as defined by \cite{Kogan2020}. In addition, exposure scores to breakthrough patents increase for routine cognitive tasks and decrease for routine manual tasks, which does not occur for 4IR patents. Aside from the general difference in methodology, \citeauthor{Kogan2020} uses an additional dimension, ``non-routine manual interpersonal tasks,'' which is not considered in our analysis and thus might impact exposure to other task types. \citeauthor{Kogan2020} conducted an analysis of patent exposure for subsets of patents, based on technology type, such as electronics patents. Future researchers can build on our patent task mapping and create similar exposure scores for a more detailed comparison of exposure scores.
The results also show changes around the year 2000, when both 4IR and non-4IR patent shares stabilized compared with previous years. Similarly, \cite{Kogan2020} observed strong changes in exposure shares before 2000 and a stabilization in the years after.

\section{Patent exposure score}

The fourth industrial revolution represents technologies which have been adopted in the current wave of technological change. These patents include, e.g., AI, machine vision, and autonomous robots. These technologies may change the way we work, even in areas that have previously been less impacted by automation technologies \citep{Brynjolfsson2017_m}.

Previous research introduced indicators on the automation potential for different occupations \citep{Frey2017,Brynjolfsson2017_m}. These indicators help to better understand which occupations and industries might be particularly impacted by the 4IR. A number of inventions play together in this progress, and diffusion might differ across occupations. However, current measures do not address the gap of technical feasibility and diffusion \citep{Arntz2019}. We address this gap by introducing a patent-based measure of exposure to 4IR technologies. Two criteria for the granting of patents includes novelty and usefulness; therefore, building on patent data should reflect technological progress \cite{Strumsky2015}.

In the following section we describe the patent exposure scores and provide some examples. Section \ref{sec:labor_impact} shows their potential use for labor market analyses, and Section \ref{sec:comparison_indicators} compares patent exposure scores with other automation and AI indicators. The exposure score data can be explored via Tableau Public\footnote{\url{https://public.tableau.com/app/profile/benjamin.meindl/viz/4IR_tech/Landing}}.

\subsection{Description of exposure scores}

We construct the 4IR exposure scores by calculating the mean of log 4IR patents per task. The 4IR exposure is higher if a task is mapped to many 4IR patents. Occupation scores are derived from the scores associated with its tasks, weighted by task importance. Therefore, occupations with many tasks, and with a high number of related 4IR patents (see Section \ref{sec:methods_4IR}), have a high 4IR exposure score. Activities such as interacting with computers and recording information have a high 4IR exposure score. Staffing organizational units and negotiating with others have the lowest 4IR exposure score. \ref{appendix:task_scores} provides more details on task exposure scores. Occupation-level 4IR exposure scores range from a value of below three to above six. The lowest 4IR exposure scores are for occupations such as Meat, Poultry \& Fish cutters/trimmers and Floor Sanders \& Finishers. High 4IR exposure scores are for occupations such as Credit Authorizers, Data Entry Keyers, Computer Network Support Specialists, and Statistical Assistants. Table \ref{tab:4IR_per_CC} provides an overview of 4IR exposure scores per SOC Career Clusters\footnote{\url{https://careertech.org/sites/default/files/Perkins_IV_Crosswalk_Table_5_SOC-ONET-Nontrad-Cluster-Pathway.xls}} from Career Technical Education (CTE) and \ref{appendix:scores} lists exposure scores per occupation.

\begin{table}[h!]
    \centering
    \caption[4IR exposure score per SOC career cluster.]{4IR exposure score per SOC career cluster.}
    \begin{tabular}{lr}
\toprule
SOC Career Clusters & Mean 4IR exposure \\
\midrule
Information Technology &          5.42 \\
Finance &          5.32 \\
Marketing &          5.17 \\
Business Management \& Administration &          4.87 \\
Government \& Public Adminstration &          4.68 \\
Science, Technology, Engineering \& Mathematics &          4.63 \\
Transportation, Distribution \& Logistics &          4.58 \\
Health Science &          4.53 \\
Arts, Audio/Video Technology \& Communications &          4.51 \\
Human Services &          4.50 \\
Education \& Training &          4.49 \\
Hospitality \& Tourism &          4.40 \\
Law, Public Safety, Corrections \& Security &          4.28 \\
Agriculture, Food \& Natural Resources &          4.17 \\
Manufacturing &          4.15 \\
Architecture \& Construction &          3.82 \\
\bottomrule
\end{tabular}

    \label{tab:4IR_per_CC}
{
\begin{flushleft}
\small
Note:  A full table of exposure scores per occupation is provided in \ref{appendix:scores}.
\end{flushleft}
}
\end{table}

On average, low-skilled occupations have a lower 4IR exposure score than medium- to high-skilled occupations, whereas exposure to overall patents is higher for lower-skilled occupations (\ref{appendix:4IR_edu} provides more information on skill levels and exposure scores). The 4IR exposure scores grew fastest during the 1990s, and growth was particularly strong for medium- and high-skilled occupations. In recent years (between 2007 and 2013), 4IR exposure grew faster for occupations with currently lower scores (thus, some occupations caught up on 4IR exposure). This trend is true, particularly for low-education occupations (less than 20\% of workers with college degrees). 
Nearly all occupations within the 50\% of occupations with the highest exposure to non-4IR patents are low-education occupations. Within 4IR exposure scores, there is less differentiation between low and high-skilled occupations, even though most of the lower quintile occupations are low-educated. This supports the findings of \cite{Brynjolfsson2017_m}, that the impact of machine learning differs from previous waves of automation, impacting many tasks previously considered not automatable.

\subsection{Patent exposure sub-scores and non-4IR exposure}

\label{sec:sub_scores}
Based on our patent occupation mapping it is possible to calculate patent exposure scores for various technology groups, as long as there is a definition available for which scores are related to which technology. Those exposure scores could include, for example, AI, or robot patents, as used by \cite{Webb2020}. For this article, we calculated sub-scores of the 4IR exposure scores. Our definition of 4IR technologies describes 367 distinct technologies (see Section \ref{sec:methods_4IR} for an overview and \citealp{Meniere2020} for detailed information on those technologies), and we provide technology exposure scores for each of the technologies. These technologies include categories such as healthcare, software, and user interfaces. Each of these categories comprises one or more technologies. A full dataset of 4IR exposure scores and sub-scores is available and \ref{appendix:4IR_subscores} provides some examples of occupation exposure to 4IR exposure sub-scores.

In addition we calculate exposure to (traditional) non-4IR patents. The analysis showed that 4IR exposure and non-4IR exposure are inversely related, and thus, that future technology impact likely differs from the past.  Construction, manufacturing, transportation, agriculture, and hospitality occupations are mainly exposed to traditional technologies. Occupations in marketing, finance, administration, education, and law, for example, have a relatively strong exposure to 4IR technologies. \ref{appendix:4IR_non_4IR} provides an overview of 4IR and non-4IR exposure per SOC Career Clusters.

\section{Impact of technologies on the labor market and patent exposure}\label{sec:labor_impact}

This section describes how the 4IR exposure scores can contribute to the debate on the impact of 4IR technologies on the labor market.
There is an increasing capacity of machines to perform tasks that previously only humans could do, and several studies predict a high risk of labor automation, such as \cite{Frey2017} who claim that 47\% of US workers are employed in occupations being ``at risk of automation" (see section \ref{sec:i40_fo}). This creates fears of machines making workers obsolete and increasing unemployment \citep{Mokyr2015,Autor2015}. 

In order to explain the impact of technologies on jobs, a leading hypothesis has been that new technologies are biased in favor of skilled workers (Skill-Biased Technological Change, SBTC).
 When the SBTC hypothesis became unable to explain all major changes in the labor market, \cite{Acemoglu2011} and 
 \cite{Dorn2013} developed a more refined explanation, the Routine Replacing Technological Change (RRTC) hypothesis.
 RRTC builds on the idea that computers are particularly efficient at performing clearly defined ``routine tasks'', and this leads to a decline in demand for human labor to perform these tasks. As the share of routine tasks is highest among middle-paid occupations, computerization was accompanied by a hollowing-out of the wage structure, with declining shares of middle-paid jobs, known as employment polarization \citep{Goos2014,Kearney2006,Oesch2011}. 
 
However, the RRTC theory does not explain fully the impact of new technologies on jobs, as automation potentials do not necessarily translate into employment losses, due to various macroeconomic adjustment processes. Using regional-level information for the US, \cite{Dorn2013}, for example, find no net negative employment effects of computerization. Also, \cite{Gregory2016a} find that computerization in Europe did not reduce employment but increased it. They show that significant replacement effects exist, which are overcompensated by productivity effects. So, the net employment effect is positive, despite large capital-labor substitution. 
 \cite{Acemoglu2018a} show in their seminal theory that the effects of new technologies crucially depend on the type of technological progress. They differentiate between different types of technological progress: (1) automation, i.e., machines learn to perform tasks which previously only humans could do, (2) deepening, i.e., machines become better at tasks already automated, and (3) reinstatement, i.e., humans take over new tasks. In addition, the diffusion of new technologies creates demand for workers who produce or maintain those technologies (capital accumulation effect). The relative size of these effects and their interaction determine the overall effect of automation on the labor market.

Patent-based exposure scores can provide a valuable indicator for technological change, and have been used in recent labor market analyses. 
\cite{Mann2017} use patent exposure per industry and identify a positive overall impact of automation patents on employment. \cite{Webb2020} found a negative impact at an occupation level.
\cite{Kogan2020} showed that recent breakthrough technologies are more related to cognitive tasks than previous waves of technological change. \cite{Acemoglu2020c} find that establishments with occupations exposed to AI patents posted more AI vacancies but fewer non-AI vacancies.

In the following section we show basic correlations and compare our patent exposure scores with other patent-based indicators to help researchers better understand the potential use of the 4IR exposure score.

\subsection{Patent exposure per wage percentile and education}

We analyze patent exposure per education level and per wage percentile to compare patent indicators with theory on SBTC and RRTC. Our analysis shows that exposure to (traditional) non-4IR patents follows expected patterns, whereas 4IR patents show different patterns. 

For the analysis of the exposure per education level we rely on two indicators. First, we show the exposure per education level of workers in occupations. Therefore we rely on O*Net data on the education level for workers per occupation. Second, we use O*Net information on Job Zones, an indicator comprising the education and experience required to work in an occupation. 
Our analysis shows that the exposure to non-4IR patents is highest for low-education skilled occupations (see \ref{appendix:4IR_edu}) and is in line with the theory of SBTC. In contrast, 4IR patent exposure is highest for medium-to-high-skilled occupations.

\begin{figure}[h]
\begin{center}
\subfigure[Exposure to Non-4IR patents]{\includegraphics[width=0.49\textwidth]{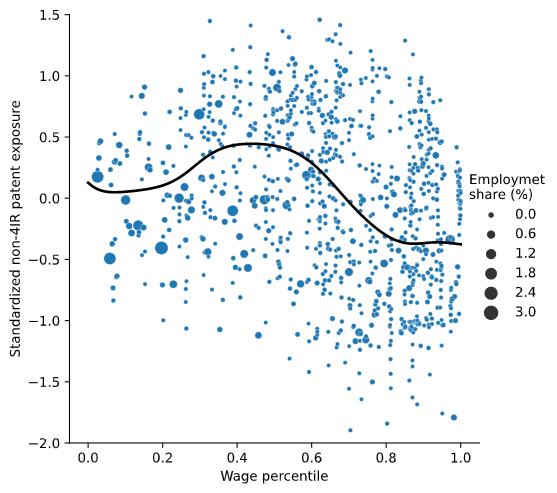}}
\subfigure[Exposure to 4IR patnets]{\includegraphics[width=0.49\textwidth]{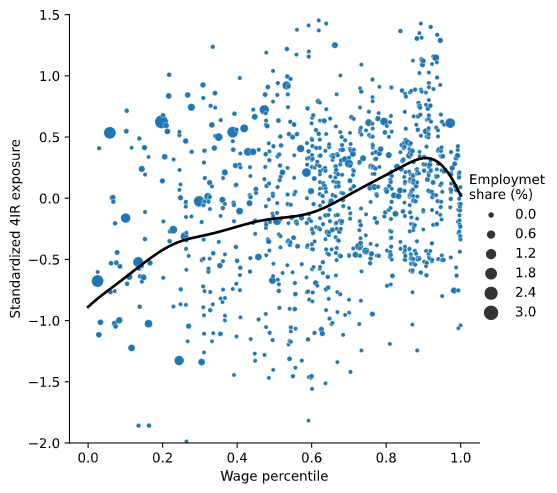}}
\caption[Exposure to patents per wage percentile.]{Exposure to patents per wage percentile. We apply a locally smoothed regression following \cite{Acemoglu2011}, using a bandwidth of 0.08 with 100 observations.}
\label{fig:exposure_per_wage}
\end{center}
\end{figure}

Next, we analyze patent exposure score per wage percentile. Therefore, we build on BLS occupation and employment statistics data\footnote{https://www.bls.gov/oes/} to extract occupation shares of workforce and wage data.
Figure \ref{fig:exposure_per_wage} shows exposure to non-4IR patents and to 4IR patents per wage percentile. 
Non-4IR exposure is highest for middle-wage occupations. This supports the ideas of RRTC, where (traditional, non-4IR) technologies mainly address routine-heavy medium-wage occupations. 
Interestingly, this observation is not confirmed at a task level. Section \ref{sec:tasktype} shows, that routine cognitive tasks have higher non-4IR exposure than non-routine cognitive tasks, but non-routine manual physical tasks have higher exposure scores than routine manual tasks.
On the one hand, this may indicate that RRTC theory is particularly appropriate for cognitive tasks, where computers have mainly been programmed to conduct routine tasks. On the other hand, this could indicate that routine intensive occupations are more likely being automated by technologies, whereas non-routine occupations are more likely to benefit from technologies, e.g., through augmentation.
The 4IR patent exposure follows a different pattern and correlates with mean income.

Overall this analysis shows different patterns of technology exposure for 4IR patents than for non-4IR patents. This suggests that the impact of 4IR technologies may differ from previous industrial revolutions, and thus confirms the findings of \cite{Brynjolfsson2017_m}, who found that future waves of automation might affect different occupations which have been considered non-automatable in the past. The exposure to wage varied for different 4IR exposure sub-scores. 4IR-manufacturing patents showed a curve similar to non-4IR patents, with highest exposure for medium-wage occupations and lower exposure for low- and high-wage occupations. 4IR-software patents showed a curve similar to the 4IR exposure curve, and 4IR-AI patents had an even stronger skew towards high-wage occupations (see \ref{appendix:wage_sub} for an overview of 4IR exposure  sub-scores exposure per wage percentile). This shows that 4IR sub-scores can be valuable for detailed assessments of the impact of technologies on occupations.

\subsection{Patent exposure and job growth} \label{sec:regression_job_growth}

In this section we calculate patent exposure scores based on 2012 task descriptions and 2012 patent data. We evaluate the relation of these exposure scores to job growth between 2012 and 2018. The analysis described in this section should provide an initial idea of potential impact patterns of 4IR patents on jobs, in order to compare 4IR exposure scores with other patent exposure scores. For more reliable insights, an analysis with micro data needs to be conducted. 

The analysis builds on BLS occupation and employment statistics data from 2012 and 2018 at an occupation level. We controlled for occupation industry shares in 2012, average education level per occupation, and mean annual wage. Section \ref{fig:mgi_4IR} showed a convex relationship of 4IR exposure and MGI estimated automation potential. Therefore, we included the quadratic value of 4IR exposure scores as an independent variable.

Time is required for an invention to have an impact on the labor market. \cite{Webb2020} argue that it can even take decades for new technologies to impact the labor market, and \cite{Kogan2020} find the largest impacts of patents on the labor market five to 20 years after patent filing. Our analysis supports these findings. We conduct an analysis with exposure scores from different years and find that coefficients were lower for more recent patent data (see \ref{appendix:historic_exposure_correlation} for regression results for different exposure scores). Patent exposure from 1992, which is 20 years before the start date of the labor market data we use, showed the highest coefficients for job growth in 2012–2018. This accounts for the time required for technologies to mature.

Regression analysis (see Table \ref{tab:corr_92}) suggests an overall negative and concave relationship of patent exposure and job growth. The overall negative relation is in line with the literature, where \cite{Kogan2020} identified a negative impact of breakthrough patents on job growth and \cite{Webb2020} identified a negative correlation of exposure to robot and software patents and job growth. Our analysis builds on occupation-level labor market data, and the results have to be confirmed with micro data analysis. In addition, the regression does not account for various effects of automation on the labor market. Therefore, even though our analysis shows a negative relation of 4IR exposure on job growth, there may be a different overall impact, considering, for example, the impact of deepening automation and capital accumulation.

\begin{table*}[h]
\centering
\caption{Exposure to 4IR patents squared and change in employment 2012-2018.}
\label{tab:corr_92}
\begin{tabular}{lp{1.3cm}p{1.3cm}p{1.3cm}p{1.3cm}p{1.3cm}p{1.3cm}}
\toprule

	&	(1)	&	(2)	&	(3)	&	(4)	&	(5)	& (6)\\
\midrule
4IR exposure\textsubscript{92}	&		&	-0.17***  \par(-4.61)	&	-0.11*** \par (-2.85)	&   -0.13*** \par (-3.00) &	-0.05 \par (-1.18)	&	-0.05 \par (-1.21)	\\
4IR exposure\textsubscript{92}$^2$	&		&		&	-0.09*** \par (-3.54)	&      &	-0.11*** \par (-4.23)	&	-0.11***  \par(-4.29)	\\
Job Zone	&	0.24***  \par(3.00)	&		&		&    0.17*** ( 2.90)   &	0.16***  \par(2.81)	&	0.21*** \par (2.67)	\\
LOG Wage	&	-0.02 \par (-0.36)	&		&		& &		&	-0.06  \par(-0.93)	\\
\midrule
Industry share	&	Yes	&	No	&	No	&	Yes	& Yes &	Yes	\\
Adjusted R$^2$	&	0.083	&	0.028	&	0.044	&  0.095     & 	0.117	&	0.117	\\
\bottomrule
\end{tabular}
\vspace{-3mm}
\begin{flushleft}
      \small
      Note: Analysis of 704 observations based on BLS labor market data from 2012 and 2018. Industry share relates to the industry share of the occupation in 2012. 4IR exposure scores are based on 1992 patent exposure per occupation. * p\textless0.10, ** p\textless0.05, *** p\textless0.01.
    \end{flushleft}
\end{table*}

\subsection{Patent exposure scores as indicators for labor market analysis}

This section discusses the potential use of our 4IR exposure scores and patent exposure scores by \cite{Webb2020} and \cite{Kogan2020}, for labor market analysis. \cite{Webb2020} divided their analysis into three parts. They calculated exposure scores for robot, software, and AI patents. \cite{Kogan2020} created an occupation-level exposure score for breakthrough patents, which differ in text content from previous patents. These categories are different from our 4IR patent exposure scores, making a direct comparison impossible.

Figure \ref{fig:4IR_Webb} indicates a slight correlation of robot and software exposure by \cite{Webb2020} to non-4IR patent exposure. Comparison with labor market indicators provides a similar overall picture.
Exposure to software, robots, and 4IR scores is inversely related to education level (see \ref{appendix:4IR_edu}) and shows a concave relationship to exposure per wage percentile, with highest scores for medium-wage occupations. The AI exposure score by Webb has, like the 4IR exposure score, highest exposure for high wage occupations. Exposure per education differs slightly: whereas AI exposure correlates with education level, 4IR exposure is concave, with highest values for medium-high education levels. 
Even though some of the results overlap and suggest that the overall mapping approaches follow similar patterns, the 4IR exposure has a different scope than the patent exposure score provided by \citeauthor{Webb2020} (see section \ref{sec:4IR_webb}). 

Similar to our approach, \cite{Kogan2020} uses text embeddings for mapping patents to occupations. Breakthrough patents have a similar exposure per wage percentile than non-4IR patents, with highest exposure for medium-wage occupations, and task-level exposure scores show similar patterns to 4IR patent exposure scores for some task types (see Section \ref{sec:tasktype} for analysis of exposure per task type). The vast majority of patents in the 1980 to 2002 period are non-4IR patents (see Figure \ref{fig:mean_pat_per_year}), so it is not surprising that the exposure per wage percentile is similar to the exposure to non-4IR patents. However, a comparison of both scores needs to be reviewed with caution, as the approaches vary in a number of aspects. On the one hand, \citeauthor{Kogan2020} present a different exposure score. They define breakthrough patents, which differ in text content from previous patents, and thus the nature of breakthrough patents changes over time. This changing scope may be appropriate for the long time frame of their analysis. Our score relates specifically to a set of selected 4IR patents. Some 4IR technologies were introduced decades ago and may no longer fall under the definition of breakthrough patents. On the other hand, \cite{Kogan2020} calculate the exposure score at an occupation level, whereas our analysis is conducted at a task level. As described in Section \ref{sec:docsim}, this leads to a number of differences, such as that the task-level mapping accounts better for the high variety of tasks conducted within an occupation.

In general, it is difficult to compare the exposure scores of the different studies, as they have a different focus. However, our analysis provides a patent occupation mapping which allows us to calculate not only 4IR exposure scores but also exposure to any other technology. Aside from the 4IR exposure and 4IR exposure sub-scores, researchers can, for example, calculate exposure to AI, software, robot, or breakthrough patents, as long as patent technology concordances are available.

Both, \cite{Webb2020} and \cite{Kogan2020} use their exposure scores to analyze the impact of technologies on the labor market. They analyze time frames from 1980 to 2010 and 1850 to 2010 respectively. They find an overall negative impact of technologies on the number of jobs. 
Their analyses are based on occupation descriptions from a given time, e.g., \cite{Webb2020} used O*NET task descriptions from 2017. While those descriptions offer accurate insights into the tasks conducted by an occupation in 2017, and thus allow us to identify the relevant patents for those tasks, they might not appropriately reflect the task descriptions of these occupations some decades ago. Therefore, the further the analysis goes back in time, the more caution is needed when reading the results.
Similarly, our analysis is based on current task descriptions by O*Net (in addition we calculate exposure scores based on 2012 task descriptions for the regression analysis). When using historic patent exposure scores, e.g., from 1992, we consider these scores valuable for explaining when technologies that shape current jobs were invented (see also Section \ref{sec:regression_job_growth}). As the patent occupation mappings are based on current task descriptions, it is possible that exposure scores do not accurately reflect patent exposure of that occupation in 1992. Therefore, we believe these exposure scores, which are based on current task descriptions, are most useful for analyzing the recent changes in the labor market. For the regression analyses of job growth 2012 to 2018 we calculated patent exposure scores based on 2012 task descriptions (our general exposure scores are based on 2020 task descriptions provided by O*Net).

\section{Direct comparison of 4IR exposure and other technology and automation indicators} \label{sec:comparison_indicators}

Section \ref{sec:labor_impact} showed the potential contribution of the 4IR exposure score for labor market analyses. In this section we compare the 4IR exposure score to other 4IR, AI, and automation indicators.
The 4IR exposure score provides a direct indicator of an occupation's exposure to technologies of the fourth industrial revolution (4IR). Whereas our patent-based 4IR exposure scores and the AI exposure by \cite{Webb2020} account for existing technological capabilities, \cite{Felten2021} provides a forward-looking indicator of occupation exposure to AI technologies based on scientific progress. 
In both cases, high exposure scores might indicate which occupations are particularly impacted by new technologies. On the one hand, this impact may include automation, which is a replacement of tasks conducted by humans, with machines. On the other hand, there is a reinstatement effect, leading to the creation of new tasks. Additionally, tasks may be augmented through new technologies.
Existing exposure scores do not differentiate between these types of change. 
There exist automation exposure scores which indicate how likely technologies are to perform tasks currently performed by humans. 
Such automation exposure scores include suitability for machine learning (SML) by \cite{Brynjolfsson2017_m}, computerization probability (CP) by \cite{Frey2017}, and automation potential by \cite{McKinsey2017}.
This section explores the relationship of the 4IR exposure score with the aforementioned scores in order to provide additional context to the 4IR exposure. Table \ref{tab:AI_indicators} provides an overview of these indicators.

\begin{landscape}

\begin{table*}[h]
\setlength{\extrarowheight}{0.2cm}
\centering
\caption{Overview of automation, AI, and 4IR indicators.}
\label{tab:AI_indicators}
\begin{tabular}{
l>{\raggedright\arraybackslash}p{2.8cm}>{\raggedright\arraybackslash}p{2.6cm}>{\raggedright\arraybackslash}p{2.6cm}>{\raggedright\arraybackslash}p{2.6cm}>{\raggedright\arraybackslash}p{2.6cm}>{\raggedright\arraybackslash}p{2.6cm}
}
\toprule
   & 4IR exposure \par (this article) & AI exposure \par \citep{Felten2021} & SML \par \citep{Brynjolfsson2017_m}  & CP \par \citep{Frey2017} & Automation potential \par \citep{McKinsey2017} & AI exposure \par \citep{Webb2020} \\
\midrule
Description 
    &   Exposure to technologies of the fourth industrial revolution (most of which enable the introduction of AI)
    &   Exposure to artificial intelligence technology
    &   Suitability for machine learning technologies
    &   Risk of being automated through AI and robotics
    &   Share of activities potentially being automated through automation technology 
    &   Exposure to AI patents \\
    
Level of analysis
    &   Tasks conducted by occupations (O*Net)
    &   Abilities required by occupations (O*Net)
    &   Tasks conducted by occupations (O*Net)
    &   Occupation level, scaled through abilities (O*Net)
    &   18 performance capabilities (based on O*Net abilities) 
    &   Tasks conducted by occupations (O*Net)\\

Index basis
    &   Relevant patents per task
    &   Scientific AI advances, as described by the EFF, per ability
    &   Online survey of SML characteristics per task
    &   Expert evaluation of selected occupations automation potential
    &   Expert evaluation of technology performance per capability &   Relevant patents per task
    \\

Measure interpretation
    &   Existing technology 4IR capabilities based on patents (current diffusion)
    &   Existing theoretical AI capabilities based on scientific articles (possibly near future diffusion)
    &   Potential future capability of ML (possible future diffusion)
    &   Potential future capabilities of technology (possible future diffusion)
    &   Potential future capabilities of technology (possible future diffusion) 
    &   Existing technology AI capabilities based on patents (current diffusion)
    \\

\bottomrule
\end{tabular}
\end{table*}
\end{landscape}

Most of the scores in our comparison refer to AI or machine learning, whereas the 4IR exposure score relates to technologies of the fourth industrial revolution. 4IR technologies, as defined in our work, are not limited to machine learning patents. Instead, they include a range of technologies, including ``core technology fields'' IT hardware, software, and connectivity technologies, as well as ``applications'' technologies, such as remote health monitoring, predictive maintenance, and smart ATMs. However, even though not all those technologies directly relate to machine learning, they allow data generation, transmission, and analysis, thus contributing to an environment that enables the application of ``smart'' AI technologies. \cite{Meindl_wp_map} explain that, with the increasing maturity of the 4IR, the concept increasingly centers around artificial intelligence. Additionally, \cite{Frank2019a} explain in their widely used framework of Industry 4.0, that several base technologies (e.g., cloud computing) support the implementation of ``smart'' (or intelligent) front end technologies. Therefore, depending on the definition of artificial intelligence, we consider that 4IR exposure scores may be comparable to the AI exposure indices. We decide not to rely on the ``core AI'' 4IR sub-exposure score (see Section \ref{sec:sub_scores} for more background on 4IR sub scores) for this comparison, which mainly relates to AI algorithms itself, but may not reflect the full potential impact of AI, e.g., through smart office solutions or autonomous vehicles.

\subsection{Comparison of 4IR exposure and AIOI index by Felten et al.}

\cite{Felten2021} create an occupation level AI exposure score (AIOE). They identify the most relevant AI capabilities based on the Electronic Frontier Foundation (EFF) AI Progress Measurement project and link those capabilities to occupational abilities, as described by O*Net. They calculate the AI exposure score as a relative measure, accounting for all abilities related to an occupation.

\begin{figure}[h!]
    \centering
    \includegraphics[width=1\textwidth]{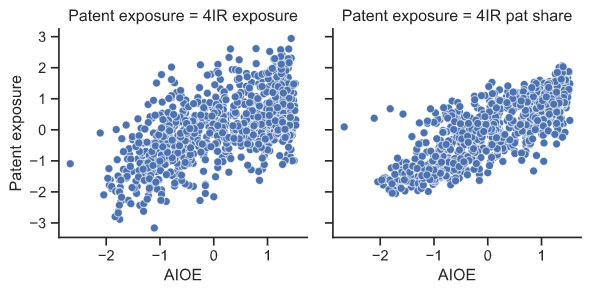}
    \caption[The graph maps the AI occupation exposure scores (AIOE) of Felten et al. to our patent exposure scores.]{The graph maps the AI occupation exposure scores of \cite{Felten2021} to our patent exposure scores. Aside from the 4IR exposure (left), we include the 4IR patent share (right), which is the share of 4IR patents to all patents. All patent exposure scores are z-scores. Both graphs show some correlation with the following correlation values: Pearson r (4IR exposure) = 0.58 and Pearson r (i40 pat share) = 0.74.}
    \label{fig:aioe_4IR}
\end{figure}

Figure \ref{fig:aioe_4IR} shows a correlation between AIOE and both 4IR exposure and the share of 4IR patents within all patents related to an occupation. The correlation is stronger for AIOE and 4IR patent share. The strong AIOE and 4IR patent share correlation is not surprising, as both measures describe occupations which mainly relate to abilities and tasks which can only be conducted using AI or 4IR technologies.
A high 4IR exposure score indicates that there is a high number of total patents related to the occupation. This could indicate that the 4IR technologies are already in a more advanced state. A high 4IR patent share does not necessarily indicate a high number of total patents, but only relates to the share of 4IR patents of all related patents. Some occupations, such as lawyers and compliance officers, have many tasks for which overall few relevant patents exist, and within those, many patents are about information and communication technologies related to the 4IR. Similarly, high AIOE occupations require many abilities which can be addressed by AI (but not by traditional technologies), independently if these technologies are already diffused. In addition, even though AI technologies have certain capabilities (such as image recognition), this does not necessarily mean that solutions have already been invented which can help workers conduct their tasks. Accordingly, \cite{Felten2021} describe their exposure score as being forward-looking.

Following this argumentation, the 4IR patent share describes the degree to which occupational tasks can be conducted or supported by 4IR technologies, whereas 4IR patent exposure takes into account the actual capabilities of current 4IR technologies.

\subsection{Combining 4IR exposure score and suitability for machine learning} \label{sec:i40_sml}

 \citeauthor{Brynjolfsson2017_m} review occupational task descriptions with regard to whether they are potentially suitable for machine learning. A high SML indicates a high risk of automation (due to the various effects described; this does not necessarily mean decreasing labor demand of related tasks). A low SML may comprise two types of tasks: first, ``old'' tasks, which are not automatable; second, ``new'' tasks, which themselves result from the introduction of machine-learning technologies such as ``Develop or apply data mining and machine learning algorithms.''

\begin{figure}[h!]
    \centering
    \includegraphics[width=1\textwidth]{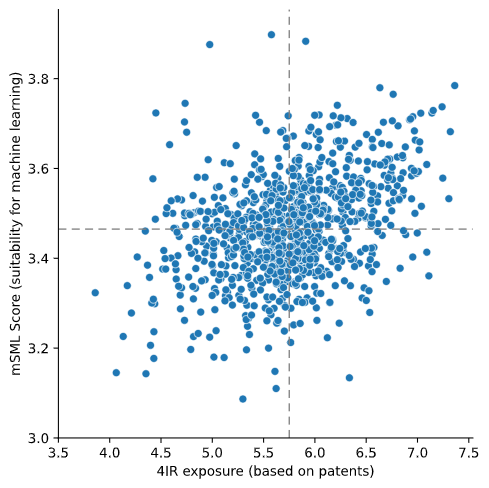}
    \caption[The graph maps 4IR exposure scores and suitability for machine learning (SML).]{The graph maps 4IR exposure scores and suitability for machine learning (SML). The graph shows a slight correlation with Pearson's r = 0.36. }
    \label{fig:sml_4IR}
\end{figure}

We consider the 4IR exposure to describe the diffusion of technologies enabling the (front end) application of machine-learning technologies. The SML score presents the general automation potential once technologies are fully implemented. Therefore, SML and 4IR exposure scores can complement each other. Figure \ref{fig:sml_4IR} maps occupations based on their 4IR exposure score and their SML score. Overall, there is a correlation, which indicates that high automation suitability is also related to a high number of 4IR patents. However, various occupations are outside this pattern; therefore we look at the graph along its four quadrants for interpretation of potential changes that occupations might undergo. We define quadrants along the mean values of the SML and the 4IR scores.

Quadrant I comprises occupations highly suitable for machine learning, and exposed to many automation patents. Through technology availability and automation suitability, these occupations might undergo significant changes in the future. They include consumer services, banking services, professional sales, web/digital communications, and travel/tourism occupations. Work activities which relate to this quadrant include ``interacting with computers,'' ``perform administrative activities,'' ``selling and influencing others,'' and ``analyzing data or information.''

Quadrant II describes occupations highly suitable for automation but with a low number of patents. These occupations might include many tasks which require more innovation before automation can happen. This group includes early childhood development and services, power, structural \& technical systems, lodging, agribusiness systems, and human resource management occupations. Work activities in this quadrant include ``judging the quality of things, services, or people''; ``establishing or maintaining interpersonal relationships''; and ``guiding, directing, and motivating subordinates.''

Quadrant III comprises occupations which might not undergo large changes through the 4IR. They only have a few tasks suitable for automation and only a few automation patents associated with those tasks. This quadrant includes construction, maintenance operations, and performing arts occupations. Work activities in this quadrant include ``updating and using relevant knowledge''; ``coaching and developing others''; and ``resolving conflicts and negotiating with others.''

Quadrant IV describes occupations exposed to a high number of 4IR patents but with many tasks with low SML. These occupations might either already have undergone changes due to the 4IR or mainly benefit from 4IR technologies through newly-created tasks or labor augmentation effects. This quadrant includes programming and software development, transportation systems or infrastructure planning, marketing communications, and professional support services occupations. Work activities in this quadrant include: ``monitoring and controlling resources''; and ``performing for or working directly with the public.''

The four quadrants could also help to describe the pressure for change to the occupations. In Quadrant I, where there is a high 4IR exposure and a high SML, there might be a high pressure on the occupations to adapt to new technologies, change tasks, reskill, or upskill. In Quadrant IV, there is also a high 4IR exposure, but due to the low SML score, there might be a lower pressure to adapt to the new technology as activities might not become obsolete as quickly, but rather benefit through augmentation. Even though there is a high SML potential for Quadrant II occupations, the pressure on those occupations is lower due to the slow diffusion of 4IR technologies in those areas. Finally, Quadrant III is impacted least by current 4IR technologies. Further analysis revealed that since 2001 the 4IR exposure grew more for occupations in the Quadrants II and III than in the highly exposed Quadrants I and IV. This delayed 4IR exposure growth suggests that technology diffusion is slower for those occupations.

Overall there is a correlation between 4IR exposure and SML scores. This correlation is strongest within the least-educated occupations (less than 10\% of workers with a college degree). The Pearson's r for low education is 0.40 vs. 0.29 for others). Low education occupations are primarily in Quadrant III and there are hardly any in Quadrant IV. High-education occupations (more than 80\% college degree) mostly have average 4IR exposure scores and are distributed across all quadrants. Medium education occupations are slightly more present in Quadrants I and IV. Therefore, there is only a small share of low-education occupations in Quadrant IV, which possibly describes occupations benefiting from 4IR technologies through labor augmentation.

Overall, the four quadrants could be described as follows in the context of the 4IR. Quadrant IV includes 4IR augmented occupations, which are likely benefiting most from 4IR occupations and have a high chance of being suitable to use labor-augmenting 4IR technologies. A relatively high share of these are highly educated. Quadrant I is currently undergoing big changes, and Quadrant II will change when more technology is developed. Quadrant III occupations have many tasks which are little impacted by the 4IR.

\subsection{Comparison of exposure scores to computerization probabilities by Frey and Osborn} \label{sec:i40_fo}

A highly-cited risk of automation index has been provided by \cite{Frey2017}. They asked a group of experts whether selected occupations may be automatable and used those ratings as a basis to calculate automation probability (CP) scores for all occupations. They consider 47\% of US automation to have a high computerization probability. \citeauthor{Frey2017} label high CP scores if their calculation leads to a risk of automation above 80\%. The results were bimodally distributed, with most occupations on the high and low extremes. \cite{Arntz2017} show that accounting for job heterogeneity within occupations leads to a more normal distribution with more medium CP jobs, and thus much fewer ``high-risk'' occupations.

We compare the CP score with SML \citep{Brynjolfsson2017_m} and find there is little or no correlation (see \ref{appendix:sml_cp} for the analysis). This is surprising, as both indices aim to estimate future automation potential for occupations. Whereas the SML score is based on a structured analysis of the suitability for machine learning at a task level, the CP has been calculated based on expert evaluations of overall occupations. Unlike SML scores, it appears that CP scores are closely related to education level.
\cite{Arntz2019} describe that this represents the idea of skill-biased technological change, which was observed before the 1980s and has been replaced by routine-replacing technological change.
One possible explanation of this bias towards low-skilled automation by \citeauthor{Frey2017} is the method they use for calculating the CP scores. They identify automation bottlenecks comprising nine skills and capabilities (features) to calculate CP scores. However, theory suggests that task content has a higher relevance for predicting automation potential than the skills a machine can perform, as skills do not directly translate into automatable tasks \citep{Acemoglu2011}. 
The SML score is calculated based on task-level indicators, and \cite{Brynjolfsson2017_m} find that artificial intelligence-driven automation might impact a much broader range of tasks and occupations than described by previous models. \citeauthor{Brynjolfsson2017_m} found that, for example, even tasks from occupations that require social interaction, such as sales and customer interaction or family and community services, are suitable for machine learning. Those occupations have low CP scores, even when they are low-educated. In turn, there are high CP scores for manual tasks such as construction, production, and maintenance, which have low or medium SML scores. Finally, CP scores are very low for most high-skilled education, whereas \cite{Brynjolfsson2017_m} find that those occupations also include a number of tasks that are suitable for machine learning. Information technology occupations, for example, have the lowest CP scores and medium SML scores.

\begin{figure}[h!]
    \centering
    \includegraphics[width=1\textwidth]{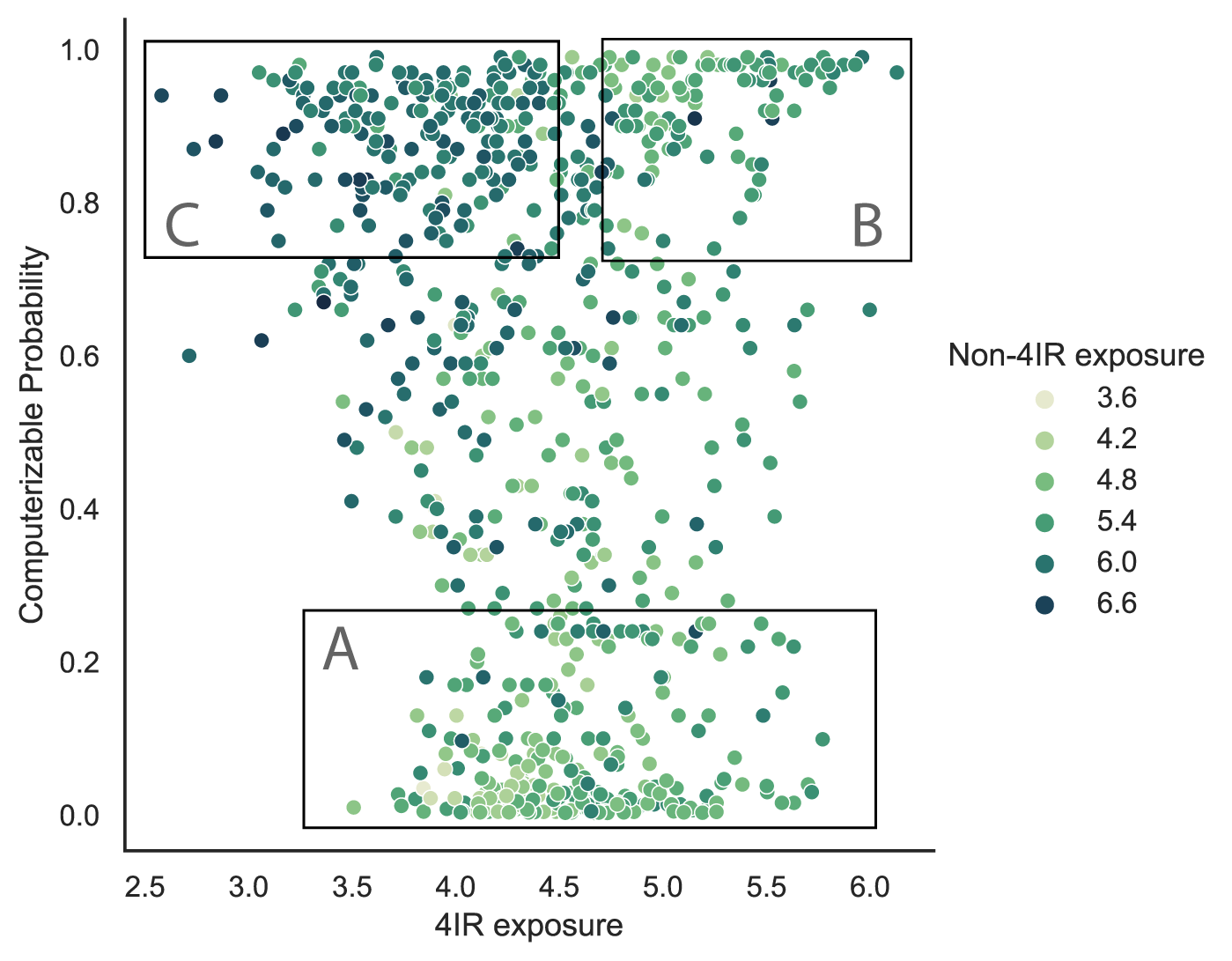}
    \caption[4IR exposure scores and Frey Osborn automation probabilities.]{4IR exposure scores and Frey Osborn automation probabilities. Section A comprises mainly high-skilled occupations. Section B comprises mainly low-skilled occupations. Section C comprises exclusively low-skilled occupations. Overall CP scores have a weak correlation with 4IR exposure (Pearson's r = -0.13) but stronger correlation to Non-4IR exposure (Pearson's r = 0.50).}
    \label{fig:fo_correlation}
\end{figure}

Next, we explore the relationship of 4IR exposure and CP scores. Figure \ref{fig:fo_correlation} shows that (similarly to the comparison of CP and SML scores) there is no overall correlation of 4IR exposure and CP. We evaluate the graph along three groups (highlighted as boxes A, B, and C) which comprise most occupations. In group A, we find that low CP occupations, which are also high-education occupations\footnote{We consider an education level high if more than 80\% of workers in an occupation hold a college degree. Results are similar if skill level is measured according to O*Net Job Zones.}, generally having low exposure to non-4IR patents. These occupations include the fields of information technology; science, technology, engineering, mathematics; education and training; human services; health sciences; and arts, video technology \& communications. Next, we describe occupations in groups B and C. Those have a high CP, are mostly low-skilled, and the 4IR exposure is inversely related to the exposure to non-4IR patents. Group B comprises those occupations with high 4IR exposure and medium non-4IR exposure, including mainly low-skilled administrative and clerical occupations in the fields of finance, marketing, and business management \& administration. Group C comprises low 4IR exposure and high non-4IR exposure occupations, which are mainly manual occupations, including the fields of transportation and logistics, hospitality \& tourism; agriculture, food \& natural resources; and manufacturing, and construction.

Overall, the two main predictors for high CP scores are low education and a medium or high exposure to non-4IR patents. While an interpretation of this observation is difficult, it raises the question of whether the expert evaluation of automation probability is biased towards skill-biased technological change and ignores the disruptive impact of machine learning, which might significantly differ from previous waves of automation.

\newpage
\subsection{Comparison of exposure scores to MGI automation scores.}

The McKinsey Global Institute \cite{McKinsey2017} conducted a detailed evaluation of the automation potential of occupations, in which they identified which sets of capabilities are required for each of over 2000 work activities (O*NET detailed work activities). For each of these capabilities they evaluated the technological readiness to automate these capabilities until 2030, which served as a basis for the evaluation of which share of jobs can be automated per occupation\footnote{The dataset can be explored through a Tableau public dataset at https://public.tableau.com/profile/mckinsey.analytics\#!/ vizhome/AutomationandUSjobs/USAutomationlandscape, accessed on April 10, 2021.}.

\begin{figure}[h]
\begin{minipage}{0.48\textwidth}
    \centering
    \includegraphics[width=0.99\textwidth]{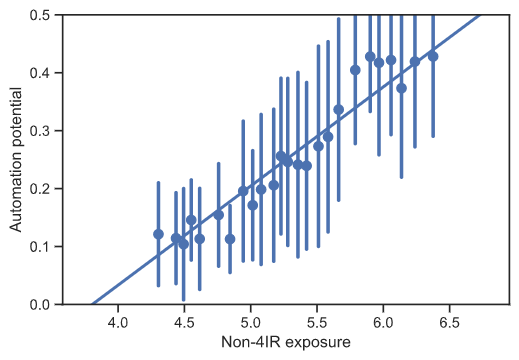}
\end{minipage}
\begin{minipage}{0.48\textwidth}
    \centering
    \includegraphics[width=0.99\textwidth]{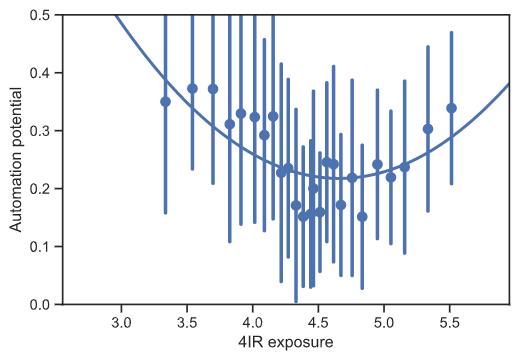}
\end{minipage}
\caption[Patent exposure score compared with MGI automation potentials.]{Patent exposure score compared with MGI automation potentials. The right graph shows that exposure to non-4IR patents is related to higher MGI automation potential (Pearson's r = 0.64). The left side shows that medium 4IR exposure relates to lowest automation potential. The graphs show data for 545 occupation in 25 bins. The vertical lines indicate the standard deviation of occupations per bin.}
\label{fig:mgi_4IR}
\end{figure}

We compare patent exposure scores with MGI automation potential estimates in Figure \ref{fig:mgi_4IR}. The correlation shows that exposure to non-4IR patents relates to higher automation potential, whereas automation potential is lowest for medium to high 4IR exposure occupations. Whereas SML and CP scores describe general automation potential, the MGI describes which automation might actually be implemented by 2030. The results suggest that most automation potential relies on non-4IR technologies, whereas occupations with higher exposure to 4IR patents (except for very high exposure) seem more prone to short-term automation. They include administrative support, sales and service, quality assurance, and health informatics occupations, and they also have high SML scores. These high 4IR and occupations are possibly the first to feel the automation impact of 4IR technologies. If we assume that high 4IR exposure is due to an advanced development of associated technologies, this is in line with \cite{Kogan2020}, who highlight that some time is required before new technologies have an impact on jobs.

\subsection{Comparison of 4IR exposure and Webb's exposure scores}\label{sec:4IR_webb}

\cite{Webb2020} created a mapping of occupations to AI, software, and robot patents. Webb uses a different NLP approach for mapping patents to jobs than the present article (for a detailed comparison see section \ref{sec:background}). Additionally, the method for construction of the exposure score differs. Whereas \cite{Webb2020} creates separate exposure scores for 3 technologies, our article provides one main exposure score to patents of the fourth industrial revolution and additionally more than 300 exposure sub-scores for a broad range of 4IR technologies, such as computer-aided design, NLP, and smart office solutions.

\begin{figure}[h]
    \centering
    \includegraphics[width=\textwidth]{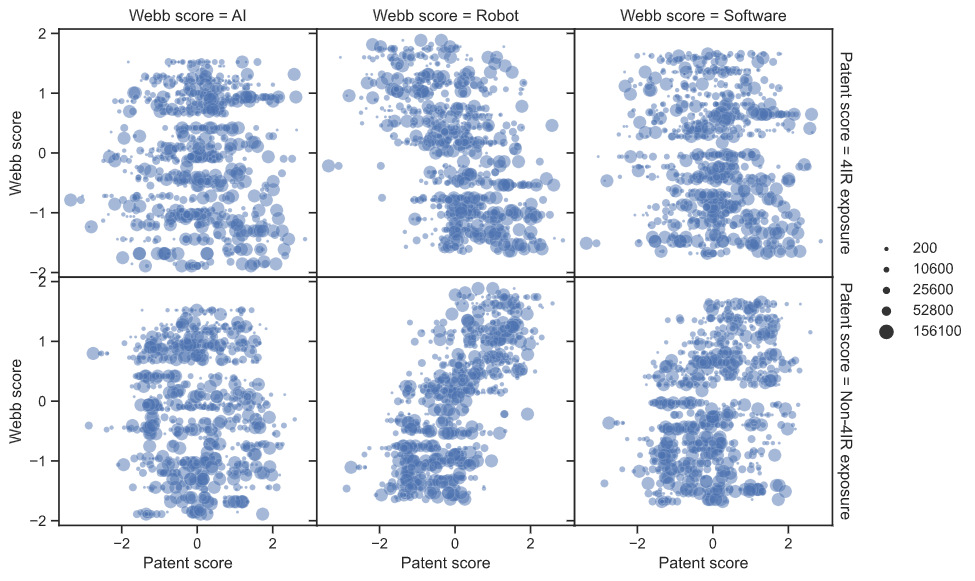}
\caption[Patent exposure scores compared with Webb exposure scores.]{Patent exposure scores compared with \cite{Webb2020} exposure scores. For comparison we use z-scores of patent exposure scores and Webb's percentile scores per technology. Bubble sizes indicate numbers of jobs. Pearson's r for the first row are 0.05, 0.44, -0.84, for the second row 0.05, 0.71, 0.42.}
\label{fig:4IR_Webb}
\end{figure}

Figure \ref{fig:4IR_Webb} compares Webb's exposure scores with our 4IR and Non-4IR patent exposure. The results show that there is no correlation for AI patents and only a slight correlation between software exposure and exposure to non-4IR patents. Also, considering 4IR sub-scores related to AI and software (see \ref{appendix:webb_sub}), we cannot find a strong correlation of Webb's scores and our 4IR exposure scores. There are various reasons for the non-correlation. First, the fundamentally different algorithm of mapping patents and tasks can be a reason why the exposure scores per occupation differ. Second, the scores created (4IR, vs. AI, software, robot) are different. Whereas our 4IR definition is very broad, Webb's definition relates to specific technologies. Further, Webb selects technology patents via text search, while our approach relies on CPC classifications of patents. Third, the aggregation method for task scores to occupation scores differs between the two approaches. 

\section{Conclusions and future work}\label{conclusions}

The aim of this paper is to better understand the exposure of occupations to technologies of the 4IR.
Several existing indicators describe the theoretical automation potential or future exposure potential of occupations.
We introduce an indicator reflecting actual technology diffusion, based on patent data.
This paper presents a method for mapping patents to tasks and introduces an occupation and task-level indicator of exposure to patents of the 4IR (4IR exposure score). We refine existing approaches to better account for task-level differences in patent exposure and the context in which an activity is conducted (e.g., diagnose \textit{machine} condition vs. diagnose \textit{patient} condition). We therefore consider that this approach offers a highly valuable contribution towards mapping patents to tasks and occupations.

Occupations with higher exposure scores may, for example, be more impacted by 4IR technologies.
The analysis shows that ratio of exposure to 4IR and non-4IR patents differs per occupation. Occupations with many manual tasks, such as manufacturing and construction, have high non-4IR exposure and low 4IR exposure, whereas many non-manual occupations, such as finance and marketing occupations, have a higher ratio of 4IR exposure.

The 4IR exposure score is also valuable as a complementary score to other technology or automation scores. 
For example, comparing theoretical and actual technology exposure can provide insights into which occupations might undergo changes through current technologies versus future diffusion. 

This direct measure of technological progress can provide highly valuable data for further exploration of the impact of technological change on employment \citep{Mitchell2017} and may serve as a source for labor market analysis to explore impact patterns of technologies on jobs.

We compared our 4IR exposure scores with labor market indicators and found that exposure to non-4IR patents is highest for medium-wage occupations, and that 4IR exposure is highest for high-wage occupations. Further, regression analysis showed a negative (concave) relation of 4IR exposure to job growth. Patent exposure 10 and 20 years ago showed higher coefficients on the impact on job growth than more recent patent exposure. The gap may reflect the time between invention and technology diffusion and is in line with findings of \cite{Kogan2020}. 
Further analysis with micro data is required to confirm these findings. To estimate the overall impact on the labor market, more complex modeling is required, e.g., considering the effect of deepening of automation or capital accumulation.

\cite{Acemoglu2019} observed that different technologies may have different impact patterns on the labor markets. Therefore, differentiating between 4IR technologies may offer additional value for labor market analyses. Researchers can build on our mapping for technology-level analysis. On the one hand, we provide technology-specific exposure scores (e.g., CAD, augmented reality for surgery, and smart office technologies). On the other hand, our mapping of patents to tasks allows researchers to build any other exposure scores, such as robots, or breakthrough patents, as long as a patent technology mapping is available.
Also, patent data is available at firm level and allows for time-varying measures.   

Our work provides an occupation (and task)-level indicator of 4IR patent exposure. Patents describe inventions, and not all inventions have an equal impact. Future work could thus further improve the indicator by accounting for a patent’s impact.
The count of patent citations is frequently discussed as potential measure for novelty and social usefulness, but its validity is ambiguous \citep{Strumsky2015}. 
Another approach is described by \cite{Kelly2018}, who describe ``breakthrough patents'' which significantly differ in text content from previous patents and thus might have particularly high impact. 

Our approach builds on occupation and task description data provided by O*Net. We take advantage of its extensive and hierarchical descriptions of occupational activities and tasks. Future work could rely on additional information provided by O*Net. For example, at a task and occupation level, the dataset indicates which technologies and tools are used, such as word processing software or programmable logic controllers. Building on this information may provide information on inventions related specifically to labor augmentation. 
The O*Net database describes occupations in the context of the US labor market. There exist concordance tables, which can help to use the patent occupation mapping in other contexts. These might provide additional accuracy to directly map patents to those regional occupation descriptions, if regional databases with similar hierarchical structures exist.

\section*{Acknowledgments}

The authors thank the Studienstiftung des Deutschen Volkes for financial support and MIT Portugal, who supported the visiting stay at the Massachusetts Institute of Technology. Further, we thank IN+ Projecto 1801P.00962.1.01 - IN+ UIDP/50009/2020 - IST-ID for financial support.

\section*{Competing interests}

We declare that we have no significant competing financial, professional, or personal interests that might have influenced the performance or presentation of the work described in this manuscript.

\bibliography{library}

\newpage
\appendix

\clearpage

\section{Patents per year in our dataset}
\begin{figure}[h]
    \centering
    \includegraphics[width=0.6\textwidth]{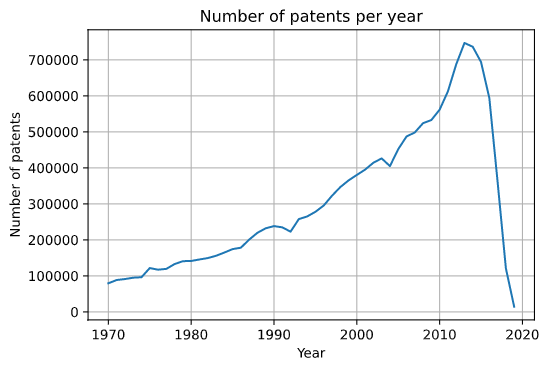}
    \caption[Number of patents in our dataset per year.]{Number of patents in our dataset per year. Our dataset includes patents with application dates between 1970 and 2019. Overall, the number of patents increased until 2013. The decline afterwards is possibly due to unpublished patents and patents not yet included in the PATSTAT dataset. Dips in patenting activity are likely due to economic decline, e.g., after the Dotcom crises in 2000. }
    \label{fig:appendix_patstat_per_year}
\end{figure}

\section{Validity check through comparing both word embeddings.} \label{appendix:validity}

Our approach compares texts of patent abstracts and task descriptions. Calculation of text similarity relies on word embedding (see Section \ref{sec:docsim}). Our results comprise relevant patents identified through two different embeddings. One embedding represents patent-specific language; the other embedding builds on a more general language (common crawl embeddings). Our approach identifies almost twice as many patents through the patent-specific embeddings. 
Only 35\% of the patent-task relations identified through the general language embeddings overlap with the patent-specific embeddings. 

We compare the results of both embeddings, to validate our overall results. 
First, we evaluate, if an embedding biases results toward certain technologies. Table \ref{cc_pt_results_tech} shows that the results per technology sector only differ slightly between both approaches. Therefore, we do not expect a bias towards certain technologies. Second, we examine whether there is a bias towards certain task types, e.g., if more patents are mapped to certain tasks. Table \ref{cc_pt_results_total} shows no indication of a bias of embeddings towards a task type. Those identical patterns indicate that the approach is robust enough so that the choice of word embedding did not systematically bias the results. 

\begin{table}[h]
\centering
\caption{The share of patents per technology type is highly similar for both word embeddings.}
\label{cc_pt_results_tech}
\begin{tabular}{lrr}
\toprule
 & \multicolumn{2}{c}{Share of patents per technology sector} \\
 & General corpus & Patent corpus \\
\midrule
Chemistry                & 8\%    & 10\% \\
Electrical engineering   & 50\%    & 46\% \\
Instruments              & 18\%    & 19\% \\
Mechanical engineering   & 16\%    & 17\% \\
Other fields             & 8\%    & 9\% \\
\bottomrule
\end{tabular}\\
{\small Note: Percentages do not sum up due to rounding}

\centering
\caption{The share of patents per task type, is highly similar for both word embeddings.}
\label{cc_pt_results_total}
\begin{tabular}{lrr}
\toprule
 & \multicolumn{2}{c}{Share of patents per task type} \\
 & General corpus & Patent corpus \\
\midrule
Information input       & 15\%    & 14\% \\
Interacting with Others & 14\%    & 16\% \\
Mental processes        & 14\%    & 12\% \\
Work output             & 58\%    & 58\% \\
\bottomrule
\end{tabular}\\
{\small Note: Percentages do not sum up due to rounding}

\end{table}

\clearpage
\section{Patent per technology fields and task type} \label{appendix:clusters}

PATSTAT provides a classification of patents into technology clusters, based on \cite{Schmoch2008}. We use this classification to evaluate our results and show to which technology clusters our results refer. These results are particularly relevant for validating our mapping.
The results show that patents in the field of IT methods for management are, on average, relevant for most tasks. Also control, computer technology, and digital communication technologies are frequently linked to occupations. Other fields, such as chemistry or nanotechnology do not provide many direct links to tasks.

\begin{figure}[h]
    \centering
    \includegraphics[width=0.9\textwidth]{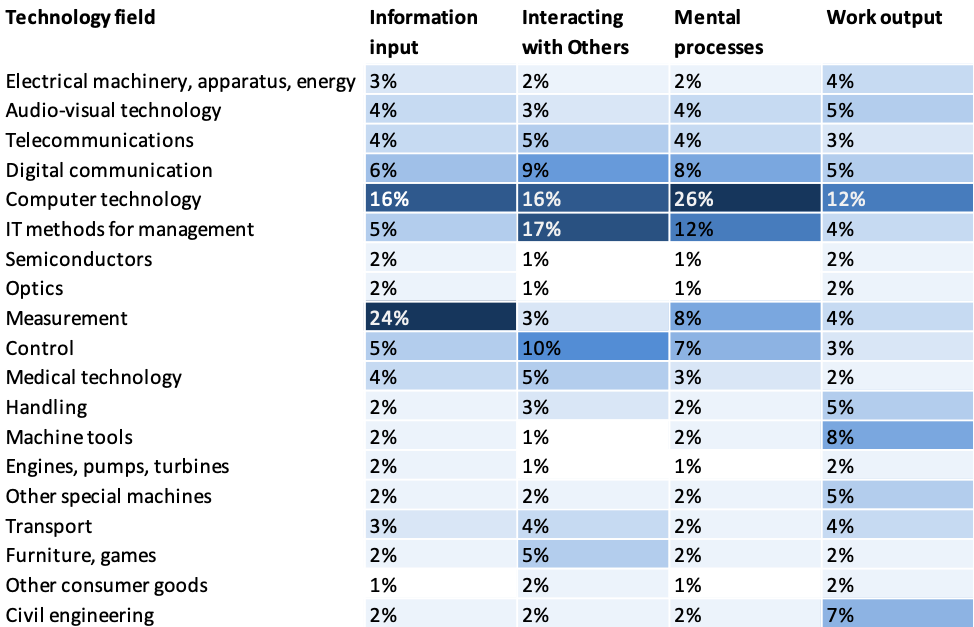}
    \caption[Technology clusters per activity group.]{Technology clusters per activity group. O*Net clusters tasks into four broad activity groups. The values indicate the shares of patents of technology clusters comprising broad activity categories. The figure includes only technology fields representing more than 1 \% of patents; thus, not all columns sum up to 100\%.}
    \label{fig:heatmap_tech_activity}
\end{figure}

Our mapping provides links of patents to occupations at a task level. In Figure \ref{fig:heatmap_tech_activity} we explore which tasks are exposed to which technology fields. Therefore, we group tasks into four high-level activity categories and evaluate the share of patents per activity category associated with each of the technology fields.

Information input tasks have a particularly high share of measurement patents; mental process tasks and interaction with other tasks have a particularly high share of computer technology, IT, and communication and control patents. Finally, work-output-related tasks have a high share of machine tool and civil engineering patents. Those findings confirm expectations on task-technology links and suggest the validity of our mapping.

\clearpage

\section{4IR exposure sub scores} \label{appendix:4IR_subscores}

\subsection{Example exposure of occupation to 4IR sub-scores}

\begin{figure}[h!]
    \centering
    \includegraphics[width=1\textwidth]{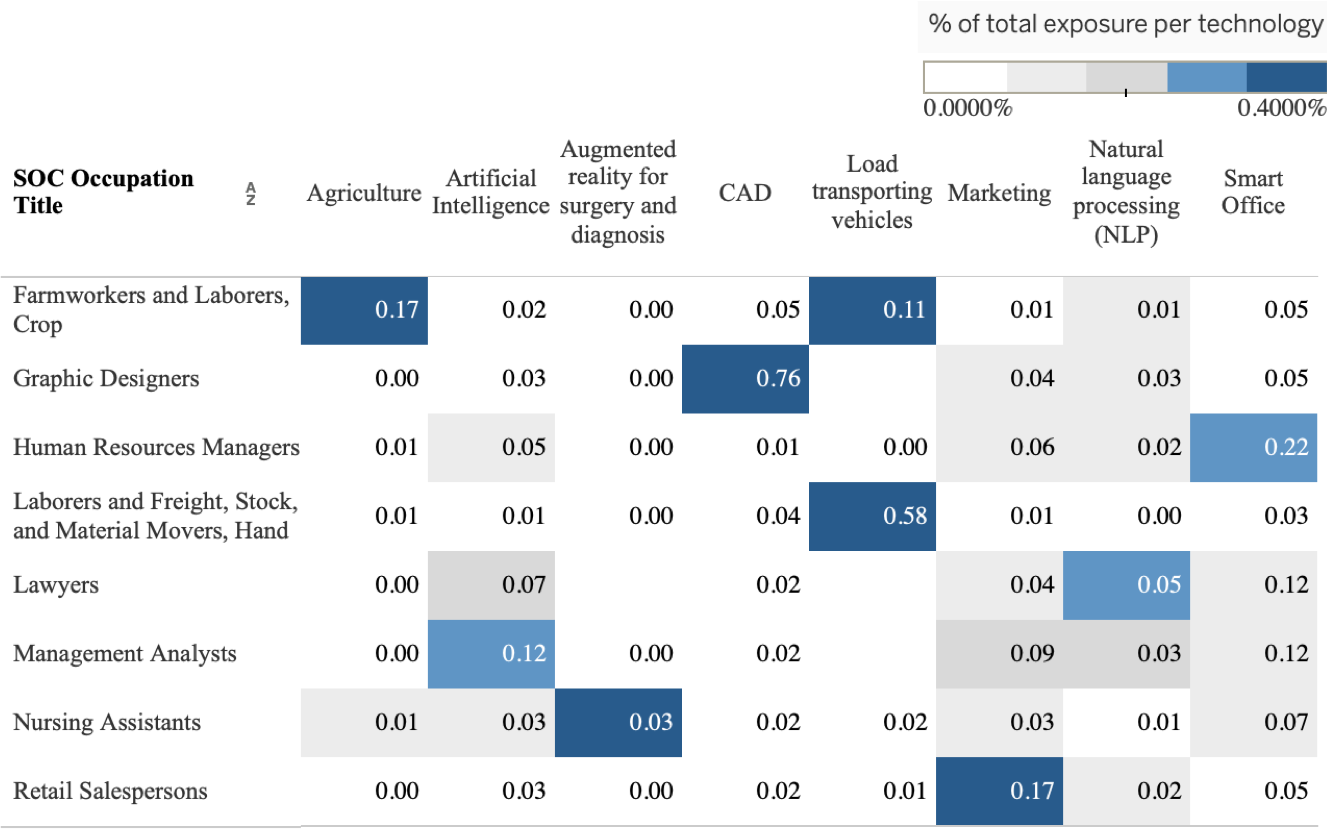}
    \caption[Example exposure of occupations to 4IR exposure sub scores.]{Example exposure of occupations to 4IR exposure sub scores. Table scores indicate the exposure of the occupation to the relevant technology. The shading refers to the share of the occupation-technology exposure of the overall exposure of the relevant technology (e.g., what share of CAD patents is related to designers).}
    \label{fig:4IR_subscores}
\end{figure}

\subsection{4IR exposure sub scores and Webb patent exposure}\label{appendix:webb_sub}

\clearpage
\section{Exposure to non-4IR patents}\label{appendix:4IR_non_4IR}
\begin{figure}[h!]
    \centering
    \includegraphics[width=1\textwidth]{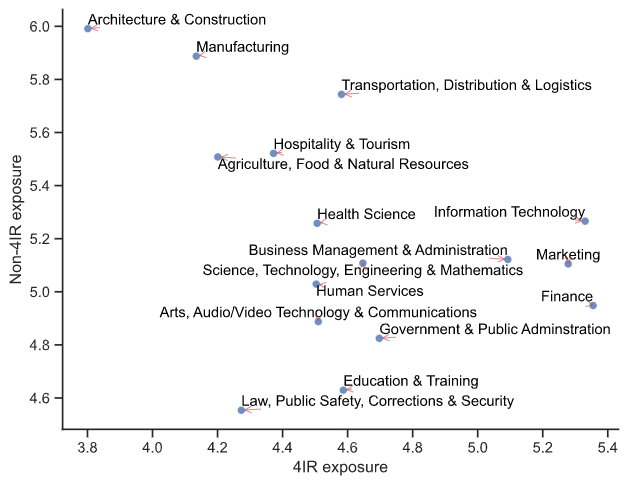}
    \caption[Exposure to 4IR vs. non-4IR patents]{Exposure to 4IR vs. non-4IR patents.}
    \label{fig:4IR_non_4IR}
\end{figure}

\clearpage

\section{Education and patent exposure}\label{appendix:4IR_edu}

\begin{figure}[h]
\begin{center}
    \subfigure[Non-4IR patents.]{\includegraphics[width=0.48\textwidth]{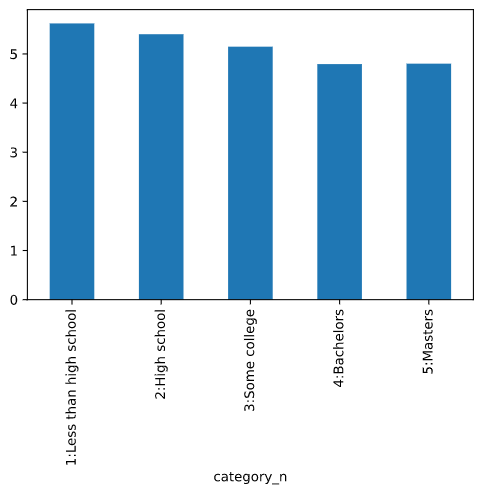}
    }
    \subfigure[4IR patents]{
    \includegraphics[width=0.48\textwidth]{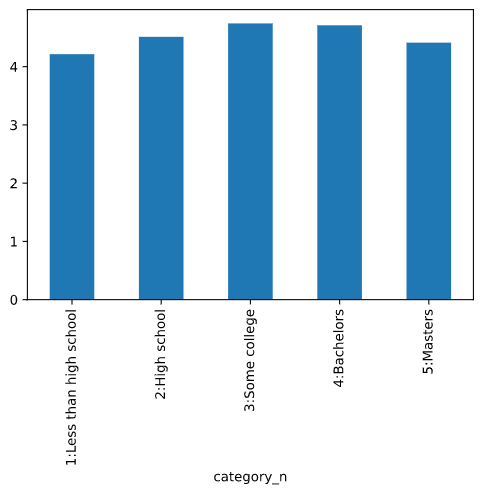}
    }
\end{center}
\caption[Exposure to patents per education level.]{Exposure to patents per education level. Education level is derived from O*Net occupation data and occupations are aggregated weighted by share of the total jobs, as provided by BLS.}
\label{fig:4IR_edu}
\end{figure}

Figure \ref{fig:4IR_edu} describes the exposure to patents per education level for non-4IR patents and 4IR patents. The analysis shows that non-4IR patents show an inverse correlation to education level, whereas 4IR exposure is highest for medium-to-high education levels. 

\cite{Webb2020} found that exposure to robot and software patents is inversely correlated to education level and is lowest for low-education occupations. This pattern is similar to our exposure scores to non-4IR patents. Webb's AI patent exposure is higher for higher-educated occupations, whereas 4IR exposure scores are highest for medium-to-high occupations.

\clearpage

\section{Impact of historic patent exposure on jobs} \label{appendix:historic_exposure_correlation}

\begin{table}[h]
\centering
\caption{Exposure to historic 4IR patents and change in employment 2012-2018.}
\label{tab:cc_pt_results_total}
\begin{tabular}{lp{1.3cm}p{1.3cm}p{1.3cm}p{1.3cm}p{1.3cm}}
\toprule
	&	(1)	&	(2)	&	(3)	&	(4)	&	(5)	\\
\hline
4IR exposure 1982	&	-0.10** (-2.00)	&		&		&		&		\\
4IR exposure 1982$^2$	&	-0.07** (-2.31)	&		&		&		&		\\
4IR exposure 1992	&		&	-0.05 (-1.18)	&		&		&		\\
4IR exposure1992$^2$	&		&	-0.11*** (-4.23)	&		&		&		\\
4IR exposure 2002	&		&		&	-0.07 (-1.42)	&		&		\\
4IR exposure 2002$^2$	&		&		&	-0.12*** (-3.76)	&		&		\\
4IR exposure 2007	&		&		&		&	-0.07 (-1.38)	&		\\
4IR exposure 2007$^2$	&		&		&		&	-0.12*** (-3.70)	&		\\
4IR exposure 2012	&		&		&		&		&	-0.07 (-1.44)	\\
4IR exposure 2012$^2$	&		&		&		&		&	-0.11*** (-3.60)	\\
\hline
Industry Education	&	Yes	&	Yes	&	Yes	&	Yes	&	Yes	\\
Adj. R$^2$	&	0.102	&	0.117	&	0.107	&	0.105	&	0.103	\\
\bottomrule
\end{tabular}

{\begin{flushleft}
      \small
      Note: Analysis of 720 observations based on BLS labor market data from 2011 and 2018. Controls IE indicates that industry and education (Job Zone) controls were included. Industry share relates to the industry share of the occupation in 2011. * p\textless0.10, ** p\textless0.05, *** p\textless0.01.
\end{flushleft}
}

\end{table}

\begin{table*}[h]
\centering
\caption{Exposure to historic 4IR patents and change in employment 2012-2018.}
\label{tab:cc_pt_results_total2}
\begin{tabular}{lp{1.3cm}p{1.3cm}p{1.3cm}p{1.3cm}p{1.3cm}}
\toprule
	&	(6)	&	(7)	&	(8)	&	(9)	&	(10)	\\
	\hline
4IR exposure 1982	&	-0.15*** (-3.18)	&		&		&		&		\\
4IR exposure 1992	&		&	-0.13*** (-3.00)	&		&		&		\\
4IR exposure 2002	&		&		&	-0.10** \par(-2.20)	&		&		\\
4IR exposure 2007	&		&		&		&	-0.10** \par(-2.03)	&		\\
4IR exposure 2012	&		&		&		&		&	-0.09* \par(-1.83)	\\
\hline
Industry Education	&	Yes	&	Yes	&	Yes	&	Yes	&	Yes	\\
Adj. R$^2$	&	0.096	&	0.095	&	0.090	&	0.089	&	0.088	\\
\bottomrule
\end{tabular}

{\begin{flushleft}
      \small
      Note: Analysis of 704 observations based on BLS labor market data from 2011 and 2018. Controls IE indicates that industry and education (Job Zone) controls were included. Industry share relates to the industry share of the occupation in 2011. * p\textless0.10, ** p\textless0.05, *** p\textless0.01.
\end{flushleft}
}

\end{table*}

\begin{table*}[h]
\centering
\caption{Exposure to 4IR patents and change in employment 2012-2018, no control variables.}
\label{tab:cc_pt_results_total3}
\begin{tabular}{lp{1.3cm}p{1.3cm}p{1.3cm}p{1.3cm}p{1.3cm}}
\toprule
	&	(11)	&	(12)	&	(13)	&	(14)	&	(15)	\\
\hline
4IR exposure 1982	&	-0.23*** (-6.33)	&		&		&		&		\\
4IR exposure 1992	&		&	-0.17*** (-4.61)	&		&		&		\\
4IR exposure 2002	&		&		&	-0.07* \par(-1.79)	&		&		\\
4IR exposure 2007	&		&		&		&	-0.05 \par(-1.35)	&		\\
4IR exposure 2012	&		&		&		&		&	-0.05 \par(-1.44)	\\
\hline
Industry Education	& No		&	No	&	No	&	No	&	No	\\
Adj. R$^2$	&	0.053	&	0.028	&	0.003	&	0.001	&	0.002	\\
\bottomrule
\end{tabular}

{\begin{flushleft}
      \small
      Note: Analysis of 720 observations based on BLS labor market data from 2011 and 2018. Controls IE indicates that industry and education (Job Zone) controls were included. Industry share relates to the industry share of the occupation in 2011. * p\textless0.10, ** p\textless0.05, *** p\textless0.01.
    \end{flushleft}
}

\end{table*}

        \clearpage
        \subsection{Comparison of SML and CP scores} \label{appendix:sml_cp}
\begin{figure}[h!]
            \centering
            \includegraphics[width=1\textwidth]{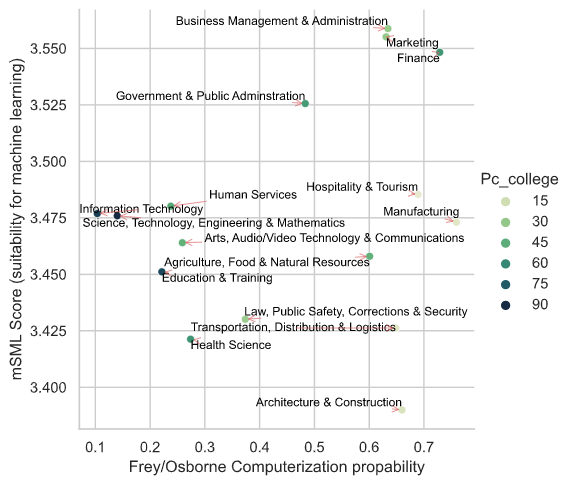}
            \caption{Comparison of SML \citep{Brynjolfsson2017_m} and CP scores \citep{Frey2017}. This graph shows a consolidated version with data grouped by SOC career clusters. The Pearson's r based on single observations is 0.078.}
            \label{fig:sml_fo_edu}
        \end{figure}

        Figure \ref{fig:sml_fo_edu} compares CP and SML scores. We group occupation scores by SOC career clusters for better readability.
        
        \clearpage
        \subsection{4IR exposure sub-scores per wage percentile}\label{appendix:wage_sub}

        \begin{figure}[h]
        \begin{center}
        \subfigure[Exposure to 4IR-Manufacturing patents]{\includegraphics[width=0.49\textwidth]{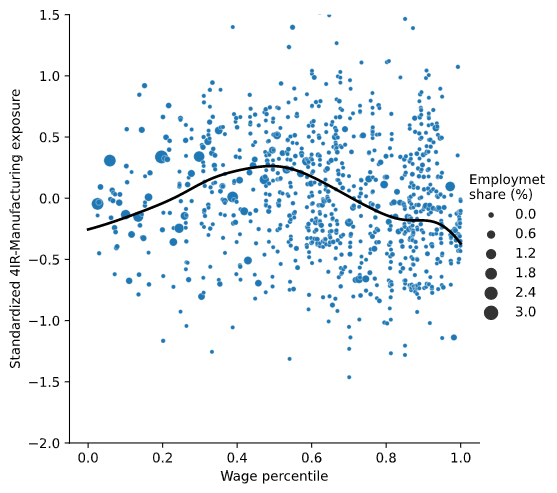}}
        \subfigure[Exposure to 4IR-Software patents]{\includegraphics[width=0.49\textwidth]{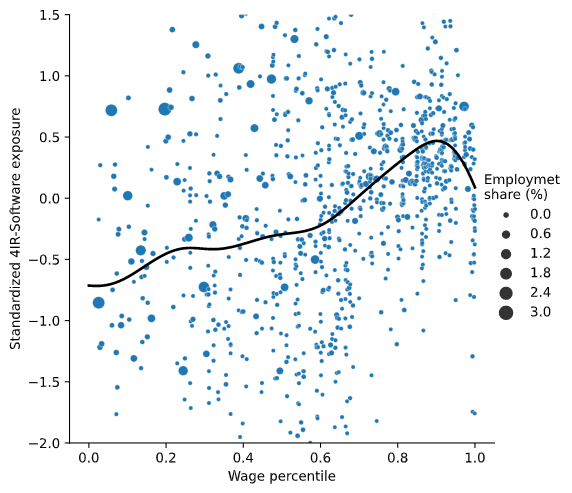}}
        \subfigure[Exposure to 4IR-AI patents]{\includegraphics[width=0.49\textwidth]{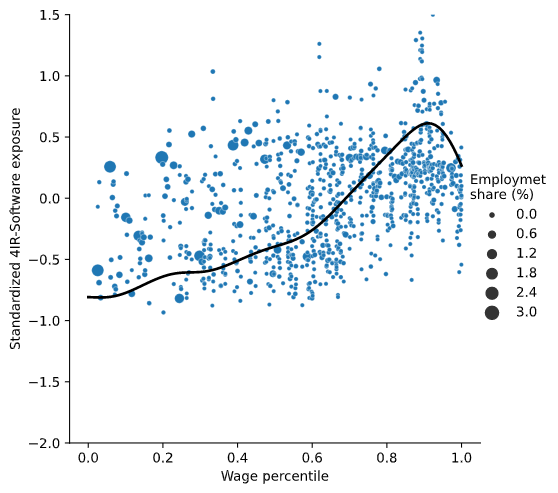}}
        \caption[4IR sub-score exposure per wage percentile.]{4IR sub-score exposure per wage percentile.}
        \label{fig:wage_4IR_sub}
        \end{center}
        \end{figure}

\clearpage
\section{4IR exposure scores}
Our approach maps patents to tasks, leading to task-level patent exposure scores. We aggregate those scores at an occupation level for occupation 4IR exposure scores. 
This section provides an overview of patent exposure scores at a task and occupation level. 

\subsection{Task- and activity-level 4IR exposure scores} \label{appendix:task_scores}

The analysis provides exposure scores for more than 20k tasks. This section provides some example scores at a task level as well as aggregated scores at a work activity level. Table \ref{tab:4IR_per_WA} shows the highest and lowest exposure scores at a work activity level.  \ref{tab:4IR_per_DWA} shows exposure scores for randomly-selected detailed work activities.

\begin{table}[h!]
    \centering
    \caption{Highest and lowest 4IR exposure scores per work activity, based on an aggregation of task-level exposure scores.}
\begin{tabular}{lr}
\toprule
                      Work Activities Element Name &  Mean 4IR exposure \\
\midrule
                        Interacting With Computers &         5.76 \\
                 Documenting/Recording Information &         5.30 \\
                     Selling or Influencing Others &         4.98 \\
                            Processing Information &         4.97 \\
                                ... & ... \\
             Controlling Machines and Processes    &         4.10 \\
    Repairing and Maintaining Mechanical Equipment &         4.05 \\
                     Staffing Organizational Units &         3.92 \\
   Resolving Conflicts and Negotiating with Others &         3.82 \\
\bottomrule
\end{tabular}

    \label{tab:4IR_per_WA}
\end{table}

\begin{table}[h!]
    \centering
    \caption{Exposure scores of randomly-selected detailed work activities (DWA).}
    \begin{tabular}{lrr}
\toprule
                                                                                  DWA Title &  4IR exposure \\
\midrule
 Develop data analysis or data management procedures. &  6.03  \\
 Measure equipment outputs. &  5.70 \\
 Correspond with customers to answer questions or resolve complaints. &  5.54  \\
 Modify software programs to improve performance. &  5.42  \\
 Manage financial activities of the organization. &  5.40 \\
 Process library materials. &  4.99 \\
 Develop plans for programs or services. &  4.77  \\
 Monitor construction operations. &  3.98  \\
 Inspect operations of green energy facilities. &  3.96  \\
 Develop organizational goals or objectives. &  3.66  \\
\bottomrule
\end{tabular}

    \label{tab:4IR_per_DWA}
\end{table}

\clearpage
\subsection{4IR exposure scores per occupation} \label{appendix:scores}

\begin{longtable}{p{.25\textwidth}p{.6\textwidth}p{.15\textwidth}p{.1\textwidth}}
\caption{Full list of 4IR patent exposure per occupation}
\endfirsthead
\endhead
\toprule
O*NET-SOC Code &                               Title &  4IR exposure \\

\midrule
43-4041.01     &                                 Credit Authorizers &          6.13 \\
43-9111.00     &                             Statistical Assistants &          6.00 \\
15-1152.00     &               Computer Network Support Specialists &          5.99 \\
43-9021.00     &                                  Data Entry Keyers &          5.96 \\
43-9041.01     &                            Insurance Claims Clerks &          5.93 \\
43-3031.00     &       Bookkeeping, Accounting, and Auditing Clerks &          5.87 \\
43-4071.00     &                                        File Clerks &          5.82 \\
\bottomrule
\end{longtable}

\end{document}